\documentclass[
reprint,
amsmath,amssymb,
aps, prl,
]{revtex4-2}

\usepackage[title]{appendix}
\usepackage{graphicx}
\usepackage{dcolumn}
\usepackage{dsfont}
\usepackage{bm}
\usepackage[hidelinks]{hyperref}
\usepackage[capitalise]{cleveref}
\usepackage{lipsum}
\usepackage[dvipsnames]{xcolor}
\usepackage{amsfonts}
\usepackage{yfonts}
\usepackage{booktabs}
\usepackage{amsthm}
\usepackage[scr=boondoxo,scrscaled=1.05]{mathalfa}

\theoremstyle{definition}

\newcommand{\rmd}{\mathrm{d}}

\newcommand{\Var}{\operatorname{Var}}

\DeclareFontFamily{U}{futm}{}
\DeclareFontShape{U}{futm}{m}{n}{
  <-> s * [.97534] fourier-bb 
  }{}
\DeclareMathAlphabet{\mathbbs}{U}{futm}{m}{n}

\begin{document}

\preprint{APS/123-QED}

\title{Kinetic and Clock Uncertainty Relations in General Non-Markovian Jump Dynamics}

\author{Jiming Zheng} \email{jiming@unc.edu}

\author{Zhiyue Lu}

\affiliation{Department of Chemistry, University of North Carolina-Chapel Hill, NC}

\date{\today}

\begin{abstract}
Accurate timekeeping underpins precision measurement and navigation and enables the temporal coordination of biological processes. At microscopic scales, time can be inferred from stochastic events, whose fluctuations impose fundamental limits on timing precision. In autonomous Markov jump processes, event order carries no additional timing information beyond transition counts when the boundary states are known. We show that memory can break this equivalence. We establish finite-time kinetic and clock uncertainty relations for jump processes with general history dependence. We identify three distinct classes of observation for time estimation and uncover a hierarchy of fundamental limits on clock precision. In an equilibrium molecular model with dynamic disorder, retaining event order eliminates the divergent contribution of slow rate fluctuations to the optimal timing variance. For thermally activated switching, this yields an exponential precision advantage over counting statistics alone. Our results establish the organization of stochastic events as a physical resource for timekeeping.
\end{abstract}

\maketitle

Accurate timekeeping is central to modern precision measurement and navigation, and the ability to measure time is essential to many biological functions. All clocks rely on physical change. From circadian rhythms and heartbeats to neuronal spikes, molecular reactions, and atomic transitions, physical processes provide signals from which elapsed time can be inferred \cite{patke2020molecular,arslanova2023perceived,shinomoto2011deciphering,ludlow2015optical}. At microscopic scales, these signals often consist of discrete stochastic events. Their random occurrence makes timekeeping inherently uncertain \cite{barato2016cost}: the same elapsed time can produce different event records, while similar records may correspond to different durations. A fundamental question in statistical physics is therefore how precisely time can be inferred from observations of stochastic dynamics.

For Markovian jump processes, fundamental precision limits have been established through thermodynamic, kinetic, and clock uncertainty relations \cite{gingrich2016dissipation,di2019kinetic,vo2022unified,pearson2021measuring,barato2016cost,blasi2026quantum,prech2025role,yan2019kinetic,macieszczak2024ultimate,prech2025optimal}. In particular, the kinetic uncertainty relation (KUR) \cite{di2019kinetic} constrains the precision of stochastic observables by the dynamical activity, or the mean number of transitions \cite{maes2020frenesy}. In the stationary long-time limit, the clock uncertainty relation (CUR) \cite{macieszczak2024ultimate,prech2025optimal} establishes a tighter precision bound for untimed event records, governed by the mean residual time from a randomly selected instant to the next transition. Together, these results establish fundamental limits on time estimation from memoryless dynamics.

However, many physical systems exhibit memory at the scale of observation. In single-molecule experiments, enzymatic turnovers exhibit long-lived correlations associated with fluctuating molecular conformations \cite{lu1998single,english2006ever,vollmar2024model}. Correlated dwell times in ion channels and history-dependent firing in neurons provide further examples \cite{rothberg1997two,truccolo2005point,farkhooi2009serial}. Such memory can emerge from Markovian microscopic dynamics when internal degrees of freedom remain unresolved \cite{hummer2015optimal,hartich2021emergent,zhao2024emergence}. Fluctuations in these hidden degrees of freedom can change the speed of the observed dynamics: more events may indicate either a longer elapsed time or a period of faster dynamics. Recent studies have shown that event order can reveal hidden transition pathways \cite{zhao2025towards} and provide information essential for thermodynamic inference \cite{blom2024milestoning,schwarz2026consistent}. Can this information also help distinguish elapsed time from fluctuations in dynamical speed? Although uncertainty relations have been established for particular classes of non-Markovian dynamics \cite{vu2020generalized,macieszczak2024ultimate,prech2025optimal,coghi2026levelprl,coghi2026level}, the role of memory in determining the fundamental limits of time estimation remains largely unexplored.

In this Letter, we establish a hierarchy of precision limits for three levels of observation of a stochastic process: unordered event counts, ordered event sequences, and complete records with timestamps. While event order provides no additional timing information beyond transition counts in autonomous Markov jump processes, we show that memory can lift this equivalence. In an equilibrium molecular model with dynamic disorder, event order reveals hidden changes in reaction speed and eliminates the divergent contribution of slow switching to the optimal timing-error rate. For thermally activated switching, this leads to an exponential precision advantage over counting alone. In a companion paper \cite{zheng2026event}, we demonstrate that retaining event order can substantially improve timing precision in experimental neuronal recordings.

Our starting point is a complete time dilation of the stochastic dynamics, which rescales the time scales associated with both transitions and memory. Using the martingale structure of stochastic trajectories, we obtain a local expression for the time-dilation Fisher information and derive finite-time kinetic and clock uncertainty relations for jump processes with general history dependence. The theory is formulated directly in terms of the conditional jump intensities of the observed dynamics, without requiring a microscopic description of the hidden degrees of freedom.

\emph{Setup and complete time dilation.---} Let us consider a non-Markovian stochastic system on a finite number of states $\Omega = \{ 1, 2, \cdots, N \}$. We label a directed jump channel by $e: j\to i$, with $i, j \in \Omega$ and $i \neq j$. Let the counting process $n_e(t)$ record the number of stochastic jumps through channel $e$, and $X_{t-}$ denotes the trajectory before $t$. Conditioned on the history $X_{t-}$, the jump probability is
\begin{equation}
  P\left[ \rmd n_e(t)=1 \,|\, X_{t-} \right] = \gamma_e^{X_{t-}}(t) \, \rmd t + o(\rmd t),
  \label{eq:conditional_rate}
\end{equation}
where $\rmd n_e(t)$ denotes the number of jumps during the infinitesimal time window $[t, t+\rmd t)$. The intensity $\gamma_e^{X_{t-}}(t)$ depends on both the current state and the entire observed history. We assume regular, non-explosive dynamics for which the probability of multiple jumps in an infinitesimal interval is $o(\rmd t)$.. The counting process obeys the Doob--Meyer decomposition \cite{meyer1962decomposition,yang2023time,stutzer2026stochastic,zheng2026mutual,zheng2026memory}
\begin{equation}
  \rmd n_e(t) = \gamma_e^{X_{t-}}(t) \, \rmd t + \rmd\varepsilon_e(t),
  \label{eq:doob_meyer}
\end{equation}
where $\rmd\varepsilon_e(t)$ is a martingale increment with zero conditional mean, $\mathbb{E}[ \rmd\varepsilon_e(t) \,|\, X_{t-} ] = 0$. Its second-order moment is
\begin{equation}
  \mathbb{E}[ \rmd\varepsilon_e(t)\rmd\varepsilon_{e'}(t') \,|\, X_{t-} ] = \delta_{ee'}\delta(t-t')\gamma_e^{X_{t-}}(t) \, \rmd t\rmd t',
  \label{eq:martingale_second_moment}
\end{equation}
which follows from the orthogonality of martingale increments and the conditional jump statistics \cite{zheng2026memory,le2016brownian}.

\begin{figure}[t]
  \centering
  \includegraphics[width=\linewidth]{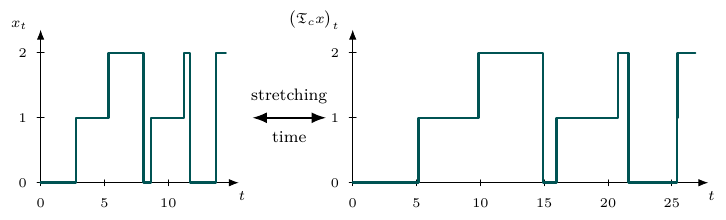}
  \caption{\textbf{Complete time dilation.} A trajectory $x_t$ of a three-state non-Markovian process (left panel) is stretched in time to give $(\mathfrak{T}_c X)(t)=x_{t/c}$ (right panel), with $c>1$. This operation is equivalent to uniformly rescaling the dynamical time scales, including those associated with memory.}
  \label{fig: time dilation}
\end{figure}

We compare the original process with a copy whose dynamical speed is rescaled by a factor $c > 0$. Let $Y$ denote the rescaled trajectory, $y_t \equiv x_{ct}$, with $x_t$ denoting system's state at $t$. The original history can be recovered by stretching $Y$ in time, as illustrated in \cref{fig: time dilation}. The stretched history and the auxiliary intensity are defined by
\begin{subequations}
\label{eq:complete_rescaling}
\begin{align}
  \left( \mathfrak{T}_cY_{t-} \right)(s) &\equiv y_{s/c}, \qquad 0\leq s<ct, \\
  \gamma_{e,c}^{Y_{t-}}(t) &\equiv c\,\gamma_e^{\mathfrak{T}_c Y_{t-}}(ct).
\end{align}
\end{subequations}
The prefactor $c$ rescales the instantaneous jump intensity, while $\mathfrak{T}_c Y_{t-}$ accounts for the simultaneous rescaling of the memory time scales. The rescaling of the memory is a purely non-Markovian effect. For an allowed transition with positive intensity, the time-dilation sensitivity is
\begin{subequations}
\label{eq:scale_sensitivity}
\begin{align}
  \left. \partial_{\ln c} \ln\gamma_{e,c}^{Y_{t-}}(t) \right|_{c=1}
  &= 1 + \left. \partial_{\ln c}
  \ln \gamma_e^{\mathfrak{T}_cY_{t-}}(ct) \right|_{c=1} \\
  &\equiv 1 + \mu_e(t).
\end{align}
\end{subequations}
where $\mu_e(t)$ denotes $\partial_{\ln c} \ln \gamma^{\mathfrak{T}_c Y_{t-}}_e \big|_{c=1}$ for convenience. The derivative is taken at a fixed observed history $Y_{t-}$. The first term is the direct speed-up of the jump intensity. The second term $\mu_e(t)$ is the additional scaling generated by stretching the memory. The term $\mu_e(t)$ is calculable once the explicit form of $\gamma_e^{X_{t-}}$ is given. For example, let $a_t$ denote the time elapsed since the most recent jump, $r_e$ a constant rate, and $\tau_m$ a fixed time scale. An age-dependent intensity $\gamma_e^{X_{t-}} = r_e(1+a_t/\tau_m)$ for an allowed transition gives $\mu_e(t) \equiv \partial_{\ln c} \ln \left[ r_e (1 + c a_t/\tau_m) \right] \big|_{c=1} = a_t/(\tau_m+a_t)$. Our general results do not require a specific functional form of $\gamma_e^{X_{t-}}$.

Let $\mathcal{P}_c^X[X_\tau]$ denote the probability density of the complete time-tagged trajectory. Differentiating its likelihood gives the score \cite{supp}
\begin{equation}
  \Lambda_\tau^X \equiv \left. \partial_{\ln c}\ln\mathcal{P}_c^X[X_\tau] \right|_{c=1} = \sum_e\int_0^\tau \left[ 1 + \mu_e(t) \right]\rmd\varepsilon_e(t).
  \label{eq:full_score}
\end{equation}
The second moment of the martingale then gives the local expression of the Fisher information \cite{supp}
\begin{equation}
  \mathcal{I}_\tau^X \equiv \left\langle \left( \Lambda_\tau^X \right)^2 \right\rangle = \sum_e \left\langle \int_0^\tau \gamma_e(t)\Big[ 1 + \mu_e(t) \Big]^2\rmd t \right\rangle.
  \label{eq:full_information}
\end{equation}
For a Markov process, $\mu_e=0$, and \cref{eq:full_information} reduces to the mean total number of jumps, i.e., dynamical activity or frenesy \cite{maes2020frenesy}. Memory can remove this reduction by making the information carried by an event depend on the history in which it occurs.

\emph{KUR, CUR, and the counting inequality.---} We consider three nested records from the non-Markovian dynamics. Let
\begin{subequations}
\label{eq:three_records}
\begin{align}
  X_\tau &\equiv \left( x_0;\{e_k,t_k\}_{k=1}^{n_\tau} \right), \\
  \Sigma_\tau &\equiv \left( x_0,e_1,e_2,\ldots,e_{n_\tau} \right), \\
  \boldsymbol N_\tau &\equiv \left( x_0,x_\tau,\{n_e(\tau)\}_e \right), \qquad n_\tau \equiv \sum_e n_e(\tau),
\end{align}
\end{subequations}
where $e_k$ and $t_k$ are the channel and timestamp of $k$th transition event. The complete record $X_\tau$ retains both the event labels and their timestamps. The sequence $\Sigma_\tau$ retains event order but discards the timestamps, while $\boldsymbol{N}_\tau$ retains only the unordered channel counts and boundary states. These three levels of observation describe the same stochastic events while retaining progressively less information. The two untimed records allow us to investigate how precisely elapsed time can be inferred without access to event timestamps.

Let $\mathcal{O}_\tau \in \{ \boldsymbol{N}_\tau, \Sigma_\tau, X_\tau \}$ be the observed records, and let $\mathcal{P}_c^{\mathcal{O}}[\mathcal{O}_\tau]$ denote its probability. Marginalizing over the discarded information gives
\begin{subequations}
\label{eq:observed_score_information}
\begin{align}
  \Lambda_\tau^{\mathcal{O}} &\equiv \left. \partial_{\ln c}\ln\mathcal{P}_c^{\mathcal{O}}[\mathcal{O}_\tau] \right|_{c=1} = \mathbb{E}\left[ \Lambda_\tau^X \,|\, \mathcal{O}_\tau \right], \\
  \mathcal{I}_\tau^{\mathcal{O}} &\equiv \langle ( \Lambda_\tau^{\mathcal{O}} )^2 \rangle.
\end{align}
\end{subequations}
Here $\Lambda_\tau^\mathcal{O}$ and $\mathcal{I}_\tau^\mathcal{O}$ are the score and Fisher information associated with the observed record $\mathcal{O}_\tau$, respectively. Consider an observable constructed from the record $\mathcal{O}_\tau$, $Q_\tau^\mathcal{O}=Q(\mathcal{O}_\tau)$. For any observable that does not depend explicitly on the waiting times or on the terminal time $\tau$, i.e., $Q_\tau^N$ and $Q_\tau^\Sigma$, complete time dilation gives \cite{supp}
\begin{equation}
  \tau\partial_\tau\left\langle Q_\tau^\mathcal{O} \right\rangle = \left\langle \left[ Q_\tau^\mathcal{O} - \left\langle Q_\tau^\mathcal{O} \right\rangle \right] \Lambda_\tau^{\mathcal{O}} \right\rangle.
  \label{eq:observed_score_identity}
\end{equation}
The Cauchy--Schwarz inequality therefore yields
\begin{equation}
  \frac{ \left[ \tau\partial_\tau\left\langle Q_\tau^\mathcal{O} \right\rangle \right]^2 }{ \Var[Q_\tau^\mathcal{O}] } \leq \langle (\Lambda_\tau^\mathcal{O})^2 \rangle = \mathcal{I}_\tau^{\mathcal{O}}.
  \label{eq:observation_level_inequality}
\end{equation}
The bound is locally optimal: an observable proportional to the accessible score, $Q_\tau^\mathcal{O} \propto \Lambda_\tau^\mathcal{O}$, reaches equality.

For a sequence observable $Q_\tau^\Sigma = Q(\Sigma_\tau)$, \cref{eq:observation_level_inequality} gives
\begin{equation}
  \frac{ \left[ \tau\partial_\tau\left\langle Q_\tau^\Sigma\right\rangle \right]^2 }{ \Var[Q_\tau^\Sigma] } \leq \mathcal{I}_\tau^\Sigma \leq \mathcal{I}_\tau^X.
  \label{eq:CUR_KUR}
\end{equation}
The rightmost bound is the non-Markovian KUR, while the sharper first bound is the non-Markovian CUR. If the observable is determined by the unordered count record, the inequality can be refined further:
\begin{equation}
  \frac{ \left[ \tau\partial_\tau\left\langle Q_\tau^N\right\rangle \right]^2 }{ \Var[Q_\tau^N] } \leq \mathcal{I}_\tau^N \leq \mathcal{I}_\tau^\Sigma \leq \mathcal{I}_\tau^X.
  \label{eq:counting_inequality}
\end{equation}
The first bound in \cref{eq:counting_inequality} is the counting inequality. Each Fisher information refers to the time-dilation information retained by the record from which the observable can be reconstructed.

The information lost when timestamps or event order are erased has an exact conditional-variance form. Orthogonality of conditional expectation gives \cite{supp}
\begin{subequations}
\label{eq:information_gaps}
\begin{align}
  \mathcal{I}_\tau^X-\mathcal{I}_\tau^\Sigma &= \mathbb{E}\left[ \Var\left( \Lambda_\tau^X \,|\, \Sigma_\tau \right) \right], \\
  \mathcal{I}_\tau^\Sigma-\mathcal{I}_\tau^N &= \mathbb{E}\left[ \Var\left( \Lambda_\tau^\Sigma \,|\, \boldsymbol N_\tau \right) \right].
\end{align}
\end{subequations}
The first gap is the time-dilation information carried by precise timestamps after the event order is known. The second is the information carried by event order after all channel counts are fixed. In a Markov process, as discussed later, the dependence of the untimed sequence probability on duration is already determined by unordered transition counts, and therefore $\mathcal{I}_\tau^\Sigma=\mathcal{I}_\tau^N$. Memory can lift this degeneracy. A strict sequence-count gap guarantees that some order-sensitive observable is more precise than any observable constructed only from unordered counts, while still obeying the CUR.

\emph{Event order information.---} For autonomous Markov jump processes, event order carries no additional timing information beyond unordered transition counts. To see this, consider a trajectory with successive waiting times $\tau_k$ and states $x_k$. Its probability density under complete time dilation is proportional to
\begin{equation}
  \mathcal{P}_c^X[X_\tau] \propto \prod_e (cr_e)^{n_e(\tau)} \exp\left[-c\sum_{k=0}^{n_\tau}\Gamma_{x_k} \tau_k\right],
  \label{eq:markov_trajectory_probability}
\end{equation}
where $r_e$ is the transition rate along edge $e$ and $\Gamma_j$ is the escape rate from state $j$. After integrating over the waiting times, the probability depends only on the number of visits to each state and the channel counts, regardless of their order. Since the unordered record $N_\tau$ completely determines these quantities, the conditional probability $P_c^\Sigma[\Sigma_\tau | N_\tau]$ is independent of $c$.
Consequently,
\begin{equation}
  \Lambda_\tau^\Sigma=\Lambda_\tau^N, \qquad \mathcal{I}_\tau^\Sigma=\mathcal{I}_\tau^N.
  \label{eq:markov_order_information}
\end{equation}
For non-Markovian dynamics, the rate of a transition generally depends on the preceding events. Reordering events may therefore change the subsequent waiting-time statistics, even when the channel counts remain unchanged. The event sequence can then carry additional timing information, giving $\mathcal{I}_\tau^\Sigma > \mathcal{I}_\tau^N$. Memory thus makes event order a distinct resource for improving clock precision.

\emph{Time estimation.---} The information bounds acquire a direct meaning as limits on clock precision. For stationary processes, define
\begin{equation}
  \mathcal{T}_\mathcal{O}^{-1} \equiv \lim_{\tau\to\infty}\frac{\mathcal{I}_\tau^\mathcal{O}}{\tau}, \qquad \mathcal{O} \in \{ X, \Sigma, \boldsymbol{N} \},
  \label{eq:information_times}
\end{equation}
provided that the long-time information rates exist. The full-record information rate retains the local expression
\begin{equation}
  \mathcal{T}_X^{-1} = \sum_e\left\langle \gamma_e(t)\Big[ 1 + \mu_e(t) \Big]^2 \right\rangle_{\rm ss}.
  \label{eq:full_information_time}
\end{equation}
The two untimed records instead determine how precisely elapsed time can be inferred without access to timestamps.

Let $Q_\tau^\mathcal{O}=Q(\mathcal{O}_\tau)$, with $\mathcal{O}_\tau \in \{\Sigma_\tau, \boldsymbol{N}_\tau\}$, be a time estimator whose readout does not depend explicitly on the unknown duration $\tau$. For a time estimator satisfying $\langle Q_\tau^\mathcal{O}\rangle - \tau = o(\sqrt{\tau})$ and $\partial_\tau \langle Q_\tau^\mathcal{O}\rangle \to 1$ as $\tau \to \infty$, the mean-square timing error and the variance have the same leading asymptotic rate. Applying \cref{eq:observation_level_inequality} gives
\begin{equation}
  \liminf_{\tau\to\infty} \frac{\Var[Q_\tau^\mathcal{O}]}{\tau} \geq \mathcal{T}_\mathcal{O}.
  \label{eq:time_estimation_precision}
\end{equation}
Thus, $\mathcal{T}_\mathcal{O}$ sets the minimum rate at which timing uncertainty can accumulate. When the bound is asymptotically attainable, the relative root-mean-square timing error approaches $\sqrt{\mathcal{T}_\mathcal{O}/\tau}$. Smaller $\mathcal{T}_\mathcal{O}$ therefore means greater timing precision.

The time scales $\mathcal{T}_\mathcal{O}$ also have a direct interpretation in terms of the observation time required to distinguish different dynamical speeds. Consider two systems whose internal time scales differ by a small relative amount $\epsilon$, corresponding to a complete time dilation $c=e^\epsilon$. The distinguishability of their observed records, measured by the Kullback--Leibler divergence, is
\begin{equation}
  D_{\rm{KL}} \! \left( \mathcal{P}_{1,\tau}^\mathcal{O} \| \mathcal{P}_{e^\epsilon,\tau}^\mathcal{O} \right) = \frac{\epsilon^2}{2} \mathcal{I}_\tau^\mathcal{O} + o(\epsilon^2) \simeq \frac{\epsilon^2\tau}{2\mathcal{T}_\mathcal{O}},
  \label{eq:time_distinguishability}
\end{equation}
in the local long-time regime. Hence, $\mathcal{T}_\mathcal{O}$ characterizes the observation time needed to resolve a given relative change in the system's dynamical speed. Let $\tau_{\rm obs}^{\mathcal{O}}(\epsilon)$ denote the observation time required to achieve a given distinguishability using the record $\mathcal{O}$. In the long-time limit,
\begin{equation}
  \lim_{\epsilon\to 0} \frac{\tau_{\rm obs}^\Sigma(\epsilon)}{\tau_{\rm obs}^N(\epsilon)} = \frac{\mathcal{T}_\Sigma}{\mathcal{T}_N}.
  \label{eq:observation_time_ratio}
\end{equation}
The information hierarchy gives $\mathcal{T}_X \leq \mathcal{T}_\Sigma \leq \mathcal{T}_N$. Event order can therefore reduce the observation time required to distinguish dynamical speeds, by exploiting information already encoded in the sequence of events.

For a stationary Markov process, the distinction between ordered and unordered records disappears. Writing $\Gamma_j$ for the escape rate from state $j$ and $\pi_j$ for the stationary probability, the time scales reduce to
\begin{subequations}
\label{eq:Markov_limit}
\begin{align}
  \mathcal{T}_X &= \mathcal{A}^{-1}, & \mathcal{A} &\equiv \sum_j \pi_j\Gamma_j, \\
  \mathcal{T}_\Sigma &= \mathcal{T}_N = \mathcal{T}_{\rm res}, & \mathcal{T}_{\rm res} &\equiv \sum_j \frac{\pi_j}{\Gamma_j} \quad \text{\cite{macieszczak2024ultimate,prech2025optimal}}.
\end{align}
\end{subequations}
Here $\mathcal{A}$ is the mean jump rate and $\mathcal{T}_{\rm res}$ is the mean time from a randomly selected stationary instant to the next jump. Thus, $\mathcal{T}_\mathcal{O}$ reduces to a characteristic waiting-time scale in Markovian dynamics, while in general non-Markovian systems its inverse measures the rate at which the retained record accumulates timing information. Memory can separate these scales, making event order useful for more precise time estimation.

\begin{figure}[htbp]
  \centering
  \includegraphics[width=1.03\linewidth]{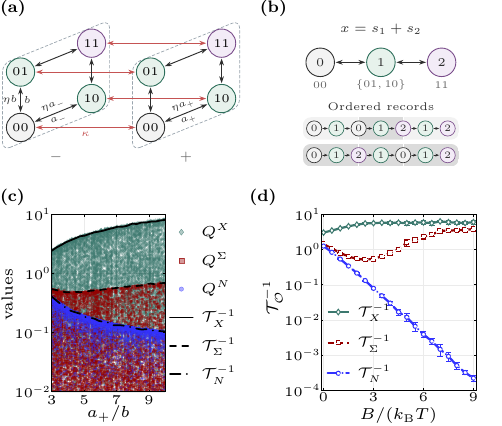}
  \caption{\textbf{Time estimation under dynamic disorder.}
    (a) Full eight-state system with microscopic states $(s_1, s_2, z)$. The two layers correspond to the catalytic modes $z \in \{ +, - \}$, which interconvert at rate $\kappa$. The gate modulates the first molecular switch through $a_z$, while the second switch is unaffected.
    (b) Coarse-grained states $x = s_1 + s_2$. The two illustrated records have identical channel counts and boundary states but different event order.
    (c) Numerical illustration of the uncertainty bounds and their hierarchy as $a_+/b$ varies. The dots represent the l.h.s. of \cref{eq:observation_level_inequality} for $Q_\tau^N$ and $Q_\tau^\Sigma$, and of \cref{eq:time_tagged_KUR} for $Q_\tau^X$, all divided by $\tau$. The lines represent the corresponding long-time information rates $\mathcal{T}_\mathcal{O}^{-1}$.
    (d) Exponential advantage of retaining event order. For thermally activated gate switching, the counting information rate $\mathcal{T}_N^{-1}$ decays exponentially with the barrier height, whereas retaining event order preserves a finite positive information rate. The long-time limit is taken at each fixed gate-switching rate before the barrier is increased. Numerical trajectories are generated using the Gillespie algorithm for the full eight-state system.}
  \label{fig: illustration}
\end{figure}

\emph{Application: Time estimation under dynamic disorder.---} Molecular reaction rates can fluctuate as internal conformations change, a phenomenon known as dynamic disorder \cite{zwanzig1990rate,10.1063/1.462993}. We consider a minimal realization in which a slowly switching catalytic gate modulates one of two reversible molecular switches. The switches have states $s_1,s_2\in\{0,1\}$, while the gate switches autonomously between $z \in \{ +, - \}$ at rate $\kappa$ in either direction. Conditional on $z$, the reactions are
\begin{equation}
  s_1:\quad 0\mathop{\rightleftarrows}^{\eta a_z}_{a_z}1, \qquad s_2:\quad 0\mathop{\rightleftarrows}^{\eta b}_{b}1.
  \label{eq:catalytic_switch_rates}
\end{equation}
The catalyst accelerates the first switch, $a_+>a_->0$, while the second switch is unaffected. Both switches have the same forward-to-reverse ratio $\eta=\exp[-\Delta g/(k_{\rm B}T)]$, so catalysis changes their reaction barriers without changing the equilibrium free-energy difference $\Delta g$. The full eight-state process (as shown in \cref{fig: illustration} (a)) satisfies detailed balance, and its equilibrium populations are independent of $a_\pm$, $b$, and $\kappa$. The observer resolves only $x=s_1+s_2\in\{0,1,2\}$, leaving the gate and the distinction between $(1,0)$ and $(0,1)$ hidden. The observed dynamics is therefore generally non-Markovian. We consider stationary trajectories and the slow-gating limit $\kappa\to0$, with $a_\pm$, $b$, and $\eta$ held fixed.

The four stationary channel fluxes at a fixed gate state are all proportional to $a_z+b$. Slow gate switching therefore produces persistent activity fluctuations along the same count direction that carries the mean timing signal. The counting information rate yields \cite{supp}
\begin{equation}
  \mathcal{T}_N
  =\frac{1}{\kappa}
  \left(\frac{a_+-a_-}{a_++a_-+2b}\right)^2+O(1).
  \label{eq:catalytic_counting_time}
\end{equation}
Thus, the counting information rate vanishes linearly with $\kappa$, even when all channel counts are retained.

Event order provides information about the catalytic mode itself. For example, with the gate frozen at rate parameter $a$, the probability that $0\to1$ is followed by $1\to2$ is
\begin{equation}
  \theta(a)
  =\frac{\eta(1+\eta)ab}
  {(a+\eta b)(b+\eta a)}.
  \label{eq:catalytic_sequence_signature}
\end{equation}
For $a_+\ne a_-$, the two modes have different sequence statistics whenever $a_+a_-\ne b^2$. A long stretch of events can then identify whether the system is in its faster or slower mode, allowing the clock readout to account for the changing reaction speed.

Let $\mathcal{T}_\Sigma^{(+)}$ and $\mathcal{T}_\Sigma^{(-)}$ be the sequence timing scales of the two reference systems obtained by freezing the catalytic rate at $a_+$ and $a_-$. A calculation based on the completed excursions $u\to1\to v$, with $u,v\in\{0,2\}$, gives
\begin{equation}
  \mathcal{T}_\Sigma
  =\frac{\mathcal{T}_\Sigma^{(+)} + \mathcal{T}_\Sigma^{(-)}}{2} + o(1), \qquad \kappa \to 0.
  \label{eq:catalytic_sequence_asymptotics}
\end{equation}
Detailed derivations and explicit forms of $\mathcal{T}_\Sigma^{(\pm)}$ are given in the Supplemental Material \cite{supp}. The factor $1/2$ reflects the stationary fractions of physical time spent in the two modes. Both reference scales, $\mathcal{T}_\Sigma^{(\pm)}$, are finite and positive. Retaining event order therefore removes the divergent contribution from slow gate switching.

For thermally activated switching, $\kappa = \kappa_0 e^{-B/(k_{\rm B}T)}$, the separation becomes exponential:
\begin{equation}
  \ln\frac{\mathcal{T}_\Sigma}{\mathcal{T}_N}
  =-\frac{B}{k_{\rm B}T}+O(1),
  \qquad\frac{B}{k_{\rm B}T}\to\infty.
  \label{eq:catalytic_exponential_advantage}
\end{equation}
Here the temperature, prefactor, and molecular rates are fixed. The limit $\tau\to\infty$ is taken at each fixed $\kappa>0$ before the barrier is increased, so the asymptotic comparison concerns $\tau\gg\kappa^{-1}$. In this regime, retaining event order reduces the optimal long-time timing variance by an exponential factor and, through \cref{eq:observation_time_ratio}, exponentially reduces the observation time required to distinguish dynamical speeds at the same distinguishability. Thus, the timing information retained by event order remains useful even as the switching time of the dynamic disorder diverges. While event order can reveal hidden dissipation in non-Markovian dynamics \cite{blom2024milestoning,schwarz2026consistent}, its value for timekeeping does not require broken detailed balance. As our equilibrium example demonstrates, event order can improve timing precision by revealing fluctuations in dynamical speed.

\emph{Conclusion and outlook.---} In this Letter, we have established finite-time kinetic and clock uncertainty relations for jump processes with general history dependence. The resulting hierarchy distinguishes the timing information available from unordered counts, ordered event sequences, and time-stamped trajectories. Memory can encode timing information in the order of events that is absent from their counts. In our equilibrium model of dynamic disorder, this information allows a stochastic clock to overcome the divergent timing variance rate caused by slow fluctuations in reaction speed, yielding an exponential precision advantage over counting alone.

These results suggest a route toward precise timekeeping that exploits fluctuations and their history rather than suppressing them. They also raise the question of how stochastic dynamics can be designed to encode useful timing information in event sequences. In our companion paper \cite{zheng2026event}, we demonstrate that such information is accessible in experimental neuronal recordings, where retaining event order reduces timing variance by up to $70\%$. More generally, our results show that the precision of a stochastic clock depends on how its events are organized as well as how frequently they occur.

\emph{Acknowledgments.---} GPT-6 Astra was used to assist with writing simulation code and polishing the manuscript. The authors take full responsibility for the work. This work is supported by the U.S. National Science Foundation under Grant No. DMR-2145256 and the Alfred P. Sloan Foundation Matter-to-Life Theory Award under Grant No. G-2025-25194.

\emph{Data availability.---} The simulation codes and numerical results supporting the findings of this Letter are openly available at \cite{data}.

\section*{End Matter}

\emph{Driven dynamics.---} The autonomous construction extends to a protocol $u_v(t)\equiv u(vt)$. To accelerate the internal dynamics while leaving the laboratory protocol unchanged, we introduce the auxiliary parameter $c$ through
\begin{equation}
  \gamma_{e,v}^{Y_{t-}}(t;c) \equiv c\,\gamma_{e,v/c}^{\mathfrak{T}_cY_{t-}}(ct),
  \label{eq:driven_rescaling}
\end{equation}
so that $u_{v/c}(ct)=u(vt)$. At $c=1$, the auxiliary intensity reduces to the original intensity $\gamma_{e,v}^{Y_{t-}}(t)$. For an observable that does not depend explicitly on waiting times, i.e., $\mathcal{O}_\tau \in \{ \Sigma_\tau, \boldsymbol{N}_\tau \}$, the score identity becomes \cite{supp}
\begin{equation}
  \left( \tau\partial_\tau-v\partial_v \right)\left\langle Q_\tau\right\rangle_v = \left\langle \left[ Q_\tau-\left\langle Q_\tau\right\rangle_v \right]\Lambda_{\tau,v}^{\mathcal{O}} \right\rangle_v,
  \label{eq:driven_score_identity}
\end{equation}
and therefore
\begin{equation}
  \frac{ \left[ \left( \tau\partial_\tau-v\partial_v \right)\left\langle Q_\tau\right\rangle_v \right]^2 }{ \Var_v[Q_\tau] } \leq \mathcal{I}_{\tau,v}^{\mathcal{O}}.
  \label{eq:driven_observation_inequality}
\end{equation}
The full score and Fisher information retain the forms in \cref{eq:full_score,eq:full_information}, with $\gamma_e$, $\mu_e$, and the ensemble average replaced by their protocol-dependent counterparts.

\emph{Time-tagged observables.---} The main text considered observables that do not depend explicitly on waiting times. For a complete trajectory with event times $0 < t_1 < \cdots < t_{n_\tau}<\tau$, define $w_0 \equiv t_1$, $w_k \equiv t_{k+1} - t_k$ for $1 \leq k < n_\tau$, and $w_{n_\tau} \equiv \tau - t_{n_\tau}$. For a trajectory with no jumps, we set $w_0 = \tau$. For a general $Q_\tau = Q(\Sigma_\tau, \{w_k\})$, changing variables under complete time dilation gives \cite{supp}
\begin{equation}
  \left. \partial_{\ln c} \left\langle Q_\tau \right\rangle_c \right|_{c=1} = \tau \partial_\tau \left\langle Q_\tau \right\rangle - \left\langle \sum_{k=0}^{n_\tau} w_k \partial_{w_k} Q_\tau \right\rangle.
  \label{eq:waiting_time_response}
\end{equation}
The corresponding full-record inequality is
\begin{equation}
  \frac{ \left[ \tau \partial_\tau \left\langle Q_\tau \right\rangle - \left\langle \sum_{k=0}^{n_\tau} w_k \partial_{w_k} Q_\tau \right\rangle \right]^2 }{ \Var[Q_\tau] } \leq \mathcal{I}_\tau^X.
  \label{eq:time_tagged_KUR}
\end{equation}
A time-tagged observable may therefore exceed the CUR bound $\mathcal{I}_\tau^\Sigma$ while remaining constrained by the full non-Markovian KUR.

\emph{Connections to self-interacting processes.---} Self-interacting jump processes provide an example in which certain forms of memory preserve the activity-based KUR \cite{coghi2026levelprl}. Consider the memory variable
\begin{equation}
  A_t^X \equiv \frac{1}{t}\int_0^t f(x_s) \,\mathrm{d}s,
  \qquad t>0,
  \label{eq:sip_state_average}
\end{equation}
where $f$ is a fixed function depending on the state $x_s$. For a channel $e: j\to i$, let the intensity be a function of $A_t^{X}$, $\gamma_e^{X_{t-}}(t) \equiv \gamma_e(A_{t-}^X)$. The dependence of $\gamma_e$ on $A$ may be nonlinear. Under complete time dilation, we obtain
\begin{equation}
  A_{ct}^{\mathfrak{T}_cY}
  = \frac{1}{ct}\int_0^{ct} f(y_{s/c}) \,\mathrm{d}s
  = A_t^Y.
  \label{eq:sip_state_invariance}
\end{equation}
Thus, the memory variable is unchanged, and the auxiliary intensity reduces to
$\gamma_{e,c}^{Y_{t-}}(t) = c\,\gamma_e^{Y_{t-}}(t)$. Consequently, $\mu_e(t)=0$ despite the history dependence of the rates.

\Cref{eq:full_information} then gives $\mathcal{I}_\tau^X = \langle n_\tau\rangle$, which is the dynamical activity. For any observable $Q_\tau$ with no explicit dependence on waiting times or on the terminal time, \cref{eq:observation_level_inequality} yields
\begin{equation}
  \frac{\left[ \tau\partial_\tau\langle Q_\tau\rangle \right]^2}
  {\operatorname{Var}[Q_\tau]}
  \leq \langle n_\tau\rangle.
  \label{eq:sip_activity_kur}
\end{equation}
This finite-time relation does not require a stationary empirical average or a separation of time scales. Under the regular long-time limits, it recovers the activity-based KUR obtained for state-observable feedback through large-deviation methods \cite{coghi2026levelprl}. Its preservation follows from the invariance of the memory under time dilation.

For feedback through more general trajectory observables, this reduction need not hold. For example, include a jump contribution in the empirical average \cite{coghi2026level},
\begin{equation}
  A_t^X
  \equiv \frac{1}{t}\int_0^t f(x_s) \,\mathrm{d}s
  + \frac{1}{t}\sum_e g_e n_e(t),
  \label{eq:sip_general_average}
\end{equation}
where $g_e$ is a fixed channel weight. Complete time dilation leaves the state contribution unchanged but multiplies the jump contribution by $c^{-1}$. For differentiable positive rates, the resulting memory sensitivity is
\begin{equation}
  \mu_e(t) = -\left[ \frac{1}{t}\sum_{e'} g_{e'} n_{e'}(t^-) \right] \left. \frac{\partial\ln \gamma_e(A)}{\partial A} \right|_{A=A_{t-}^X}.
  \label{eq:sip_memory_sensitivity}
\end{equation}
This contribution generally does not vanish. The full-record information therefore need not equal the activity, and an activity-based KUR is no longer guaranteed \cite{coghi2026level}.

The general non-Markovian KUR derived in this Letter nevertheless remains applicable, with the full-record information given by \cref{eq:full_information}. Marginalizing timestamps and event order also preserves the hierarchy in \cref{eq:CUR_KUR,eq:counting_inequality}. Even when $\mu_e=0$, the activity reduction does not imply $\mathcal{I}_\tau^N = \mathcal{I}_\tau^\Sigma$. Thus, scale-invariant feedback preserves the activity-based KUR, while our information bounds and their hierarchy remain valid without this restriction.

\bibliography{manuscript}

\begin{thebibliography}{41}%
\makeatletter
\providecommand \@ifxundefined [1]{%
 \@ifx{#1\undefined}
}%
\providecommand \@ifnum [1]{%
 \ifnum #1\expandafter \@firstoftwo
 \else \expandafter \@secondoftwo
 \fi
}%
\providecommand \@ifx [1]{%
 \ifx #1\expandafter \@firstoftwo
 \else \expandafter \@secondoftwo
 \fi
}%
\providecommand \natexlab [1]{#1}%
\providecommand \enquote  [1]{``#1''}%
\providecommand \bibnamefont  [1]{#1}%
\providecommand \bibfnamefont [1]{#1}%
\providecommand \citenamefont [1]{#1}%
\providecommand \href@noop [0]{\@secondoftwo}%
\providecommand \href [0]{\begingroup \@sanitize@url \@href}%
\providecommand \@href[1]{\@@startlink{#1}\@@href}%
\providecommand \@@href[1]{\endgroup#1\@@endlink}%
\providecommand \@sanitize@url [0]{\catcode `\\12\catcode `\$12\catcode
  `\&12\catcode `\#12\catcode `\^12\catcode `\_12\catcode `\%12\relax}%
\providecommand \@@startlink[1]{}%
\providecommand \@@endlink[0]{}%
\providecommand \url  [0]{\begingroup\@sanitize@url \@url }%
\providecommand \@url [1]{\endgroup\@href {#1}{\urlprefix }}%
\providecommand \urlprefix  [0]{URL }%
\providecommand \Eprint [0]{\href }%
\providecommand \doibase [0]{https://doi.org/}%
\providecommand \selectlanguage [0]{\@gobble}%
\providecommand \bibinfo  [0]{\@secondoftwo}%
\providecommand \bibfield  [0]{\@secondoftwo}%
\providecommand \translation [1]{[#1]}%
\providecommand \BibitemOpen [0]{}%
\providecommand \bibitemStop [0]{}%
\providecommand \bibitemNoStop [0]{.\EOS\space}%
\providecommand \EOS [0]{\spacefactor3000\relax}%
\providecommand \BibitemShut  [1]{\csname bibitem#1\endcsname}%
\let\auto@bib@innerbib\@empty
\bibitem [{\citenamefont {Patke}\ \emph {et~al.}(2020)\citenamefont {Patke},
  \citenamefont {Young},\ and\ \citenamefont {Axelrod}}]{patke2020molecular}%
  \BibitemOpen
  \bibfield  {author} {\bibinfo {author} {\bibfnamefont {A.}~\bibnamefont
  {Patke}}, \bibinfo {author} {\bibfnamefont {M.~W.}\ \bibnamefont {Young}},\
  and\ \bibinfo {author} {\bibfnamefont {S.}~\bibnamefont {Axelrod}},\
  }\bibfield  {title} {\bibinfo {title} {Molecular mechanisms and physiological
  importance of circadian rhythms},\ }\href@noop {} {\bibfield  {journal}
  {\bibinfo  {journal} {Nature reviews Molecular cell biology}\ }\textbf
  {\bibinfo {volume} {21}},\ \bibinfo {pages} {67} (\bibinfo {year}
  {2020})}\BibitemShut {NoStop}%
\bibitem [{\citenamefont {Arslanova}\ \emph {et~al.}(2023)\citenamefont
  {Arslanova}, \citenamefont {Kotsaris},\ and\ \citenamefont
  {Tsakiris}}]{arslanova2023perceived}%
  \BibitemOpen
  \bibfield  {author} {\bibinfo {author} {\bibfnamefont {I.}~\bibnamefont
  {Arslanova}}, \bibinfo {author} {\bibfnamefont {V.}~\bibnamefont
  {Kotsaris}},\ and\ \bibinfo {author} {\bibfnamefont {M.}~\bibnamefont
  {Tsakiris}},\ }\bibfield  {title} {\bibinfo {title} {Perceived time expands
  and contracts within each heartbeat},\ }\href@noop {} {\bibfield  {journal}
  {\bibinfo  {journal} {Current biology}\ }\textbf {\bibinfo {volume} {33}},\
  \bibinfo {pages} {1389} (\bibinfo {year} {2023})}\BibitemShut {NoStop}%
\bibitem [{\citenamefont {Shinomoto}\ \emph {et~al.}(2011)\citenamefont
  {Shinomoto}, \citenamefont {Omi}, \citenamefont {Mita}, \citenamefont
  {Mushiake}, \citenamefont {Shima}, \citenamefont {Matsuzaka},\ and\
  \citenamefont {Tanji}}]{shinomoto2011deciphering}%
  \BibitemOpen
  \bibfield  {author} {\bibinfo {author} {\bibfnamefont {S.}~\bibnamefont
  {Shinomoto}}, \bibinfo {author} {\bibfnamefont {T.}~\bibnamefont {Omi}},
  \bibinfo {author} {\bibfnamefont {A.}~\bibnamefont {Mita}}, \bibinfo {author}
  {\bibfnamefont {H.}~\bibnamefont {Mushiake}}, \bibinfo {author}
  {\bibfnamefont {K.}~\bibnamefont {Shima}}, \bibinfo {author} {\bibfnamefont
  {Y.}~\bibnamefont {Matsuzaka}},\ and\ \bibinfo {author} {\bibfnamefont
  {J.}~\bibnamefont {Tanji}},\ }\bibfield  {title} {\bibinfo {title}
  {Deciphering elapsed time and predicting action timing from neuronal
  population signals},\ }\href@noop {} {\bibfield  {journal} {\bibinfo
  {journal} {Frontiers in computational neuroscience}\ }\textbf {\bibinfo
  {volume} {5}},\ \bibinfo {pages} {29} (\bibinfo {year} {2011})}\BibitemShut
  {NoStop}%
\bibitem [{\citenamefont {Ludlow}\ \emph {et~al.}(2015)\citenamefont {Ludlow},
  \citenamefont {Boyd}, \citenamefont {Ye}, \citenamefont {Peik},\ and\
  \citenamefont {Schmidt}}]{ludlow2015optical}%
  \BibitemOpen
  \bibfield  {author} {\bibinfo {author} {\bibfnamefont {A.~D.}\ \bibnamefont
  {Ludlow}}, \bibinfo {author} {\bibfnamefont {M.~M.}\ \bibnamefont {Boyd}},
  \bibinfo {author} {\bibfnamefont {J.}~\bibnamefont {Ye}}, \bibinfo {author}
  {\bibfnamefont {E.}~\bibnamefont {Peik}},\ and\ \bibinfo {author}
  {\bibfnamefont {P.~O.}\ \bibnamefont {Schmidt}},\ }\bibfield  {title}
  {\bibinfo {title} {Optical atomic clocks},\ }\href@noop {} {\bibfield
  {journal} {\bibinfo  {journal} {Reviews of Modern Physics}\ }\textbf
  {\bibinfo {volume} {87}},\ \bibinfo {pages} {637} (\bibinfo {year}
  {2015})}\BibitemShut {NoStop}%
\bibitem [{\citenamefont {Barato}\ and\ \citenamefont
  {Seifert}(2016)}]{barato2016cost}%
  \BibitemOpen
  \bibfield  {author} {\bibinfo {author} {\bibfnamefont {A.~C.}\ \bibnamefont
  {Barato}}\ and\ \bibinfo {author} {\bibfnamefont {U.}~\bibnamefont
  {Seifert}},\ }\bibfield  {title} {\bibinfo {title} {Cost and precision of
  brownian clocks},\ }\href@noop {} {\bibfield  {journal} {\bibinfo  {journal}
  {Physical Review X}\ }\textbf {\bibinfo {volume} {6}},\ \bibinfo {pages}
  {041053} (\bibinfo {year} {2016})}\BibitemShut {NoStop}%
\bibitem [{\citenamefont {Gingrich}\ \emph {et~al.}(2016)\citenamefont
  {Gingrich}, \citenamefont {Horowitz}, \citenamefont {Perunov},\ and\
  \citenamefont {England}}]{gingrich2016dissipation}%
  \BibitemOpen
  \bibfield  {author} {\bibinfo {author} {\bibfnamefont {T.~R.}\ \bibnamefont
  {Gingrich}}, \bibinfo {author} {\bibfnamefont {J.~M.}\ \bibnamefont
  {Horowitz}}, \bibinfo {author} {\bibfnamefont {N.}~\bibnamefont {Perunov}},\
  and\ \bibinfo {author} {\bibfnamefont {J.~L.}\ \bibnamefont {England}},\
  }\bibfield  {title} {\bibinfo {title} {Dissipation bounds all steady-state
  current fluctuations},\ }\href@noop {} {\bibfield  {journal} {\bibinfo
  {journal} {Physical review letters}\ }\textbf {\bibinfo {volume} {116}},\
  \bibinfo {pages} {120601} (\bibinfo {year} {2016})}\BibitemShut {NoStop}%
\bibitem [{\citenamefont {Di~Terlizzi}\ and\ \citenamefont
  {Baiesi}(2019)}]{di2019kinetic}%
  \BibitemOpen
  \bibfield  {author} {\bibinfo {author} {\bibfnamefont {I.}~\bibnamefont
  {Di~Terlizzi}}\ and\ \bibinfo {author} {\bibfnamefont {M.}~\bibnamefont
  {Baiesi}},\ }\bibfield  {title} {\bibinfo {title} {Kinetic uncertainty
  relation},\ }\href@noop {} {\bibfield  {journal} {\bibinfo  {journal}
  {Journal of Physics A: Mathematical and Theoretical}\ }\textbf {\bibinfo
  {volume} {52}},\ \bibinfo {pages} {02LT03} (\bibinfo {year}
  {2019})}\BibitemShut {NoStop}%
\bibitem [{\citenamefont {Vo}\ \emph {et~al.}(2022)\citenamefont {Vo},
  \citenamefont {Van~Vu},\ and\ \citenamefont {Hasegawa}}]{vo2022unified}%
  \BibitemOpen
  \bibfield  {author} {\bibinfo {author} {\bibfnamefont {V.~T.}\ \bibnamefont
  {Vo}}, \bibinfo {author} {\bibfnamefont {T.}~\bibnamefont {Van~Vu}},\ and\
  \bibinfo {author} {\bibfnamefont {Y.}~\bibnamefont {Hasegawa}},\ }\bibfield
  {title} {\bibinfo {title} {Unified thermodynamic--kinetic uncertainty
  relation},\ }\href@noop {} {\bibfield  {journal} {\bibinfo  {journal}
  {Journal of Physics A: Mathematical and Theoretical}\ }\textbf {\bibinfo
  {volume} {55}},\ \bibinfo {pages} {405004} (\bibinfo {year}
  {2022})}\BibitemShut {NoStop}%
\bibitem [{\citenamefont {Pearson}\ \emph {et~al.}(2021)\citenamefont
  {Pearson}, \citenamefont {Guryanova}, \citenamefont {Erker}, \citenamefont
  {Laird}, \citenamefont {Briggs}, \citenamefont {Huber},\ and\ \citenamefont
  {Ares}}]{pearson2021measuring}%
  \BibitemOpen
  \bibfield  {author} {\bibinfo {author} {\bibfnamefont {A.~N.}\ \bibnamefont
  {Pearson}}, \bibinfo {author} {\bibfnamefont {Y.}~\bibnamefont {Guryanova}},
  \bibinfo {author} {\bibfnamefont {P.}~\bibnamefont {Erker}}, \bibinfo
  {author} {\bibfnamefont {E.~A.}\ \bibnamefont {Laird}}, \bibinfo {author}
  {\bibfnamefont {G.~A.~D.}\ \bibnamefont {Briggs}}, \bibinfo {author}
  {\bibfnamefont {M.}~\bibnamefont {Huber}},\ and\ \bibinfo {author}
  {\bibfnamefont {N.}~\bibnamefont {Ares}},\ }\bibfield  {title} {\bibinfo
  {title} {Measuring the thermodynamic cost of timekeeping},\ }\href@noop {}
  {\bibfield  {journal} {\bibinfo  {journal} {Physical Review X}\ }\textbf
  {\bibinfo {volume} {11}},\ \bibinfo {pages} {021029} (\bibinfo {year}
  {2021})}\BibitemShut {NoStop}%
\bibitem [{\citenamefont {Blasi}\ \emph {et~al.}(2026)\citenamefont {Blasi},
  \citenamefont {Rodr{\'\i}guez}, \citenamefont {Moskalets}, \citenamefont
  {L{\'o}pez},\ and\ \citenamefont {Haack}}]{blasi2026quantum}%
  \BibitemOpen
  \bibfield  {author} {\bibinfo {author} {\bibfnamefont {G.}~\bibnamefont
  {Blasi}}, \bibinfo {author} {\bibfnamefont {R.~R.}\ \bibnamefont
  {Rodr{\'\i}guez}}, \bibinfo {author} {\bibfnamefont {M.}~\bibnamefont
  {Moskalets}}, \bibinfo {author} {\bibfnamefont {R.}~\bibnamefont
  {L{\'o}pez}},\ and\ \bibinfo {author} {\bibfnamefont {G.}~\bibnamefont
  {Haack}},\ }\bibfield  {title} {\bibinfo {title} {Quantum kinetic uncertainty
  relations in mesoscopic conductors at strong coupling},\ }\href@noop {}
  {\bibfield  {journal} {\bibinfo  {journal} {Physical Review Letters}\
  }\textbf {\bibinfo {volume} {137}},\ \bibinfo {pages} {056302} (\bibinfo
  {year} {2026})}\BibitemShut {NoStop}%
\bibitem [{\citenamefont {Prech}\ \emph
  {et~al.}(2025{\natexlab{a}})\citenamefont {Prech}, \citenamefont {Potts},\
  and\ \citenamefont {Landi}}]{prech2025role}%
  \BibitemOpen
  \bibfield  {author} {\bibinfo {author} {\bibfnamefont {K.}~\bibnamefont
  {Prech}}, \bibinfo {author} {\bibfnamefont {P.~P.}\ \bibnamefont {Potts}},\
  and\ \bibinfo {author} {\bibfnamefont {G.~T.}\ \bibnamefont {Landi}},\
  }\bibfield  {title} {\bibinfo {title} {Role of quantum coherence in kinetic
  uncertainty relations},\ }\href@noop {} {\bibfield  {journal} {\bibinfo
  {journal} {Physical Review Letters}\ }\textbf {\bibinfo {volume} {134}},\
  \bibinfo {pages} {020401} (\bibinfo {year} {2025}{\natexlab{a}})}\BibitemShut
  {NoStop}%
\bibitem [{\citenamefont {Yan}\ \emph {et~al.}(2019)\citenamefont {Yan},
  \citenamefont {Hilfinger}, \citenamefont {Vinnicombe},\ and\ \citenamefont
  {Paulsson}}]{yan2019kinetic}%
  \BibitemOpen
  \bibfield  {author} {\bibinfo {author} {\bibfnamefont {J.}~\bibnamefont
  {Yan}}, \bibinfo {author} {\bibfnamefont {A.}~\bibnamefont {Hilfinger}},
  \bibinfo {author} {\bibfnamefont {G.}~\bibnamefont {Vinnicombe}},\ and\
  \bibinfo {author} {\bibfnamefont {J.}~\bibnamefont {Paulsson}},\ }\bibfield
  {title} {\bibinfo {title} {Kinetic uncertainty relations for the control of
  stochastic reaction networks},\ }\href@noop {} {\bibfield  {journal}
  {\bibinfo  {journal} {Physical review letters}\ }\textbf {\bibinfo {volume}
  {123}},\ \bibinfo {pages} {108101} (\bibinfo {year} {2019})}\BibitemShut
  {NoStop}%
\bibitem [{\citenamefont {Macieszczak}(2024)}]{macieszczak2024ultimate}%
  \BibitemOpen
  \bibfield  {author} {\bibinfo {author} {\bibfnamefont {K.}~\bibnamefont
  {Macieszczak}},\ }\bibfield  {title} {\bibinfo {title} {Ultimate kinetic
  uncertainty relation and optimal performance of stochastic clocks},\
  }\href@noop {} {\bibfield  {journal} {\bibinfo  {journal} {arXiv preprint
  arXiv:2407.09839}\ } (\bibinfo {year} {2024})}\BibitemShut {NoStop}%
\bibitem [{\citenamefont {Prech}\ \emph
  {et~al.}(2025{\natexlab{b}})\citenamefont {Prech}, \citenamefont {Landi},
  \citenamefont {Meier}, \citenamefont {Nurgalieva}, \citenamefont {Potts},
  \citenamefont {Silva},\ and\ \citenamefont {Mitchison}}]{prech2025optimal}%
  \BibitemOpen
  \bibfield  {author} {\bibinfo {author} {\bibfnamefont {K.}~\bibnamefont
  {Prech}}, \bibinfo {author} {\bibfnamefont {G.~T.}\ \bibnamefont {Landi}},
  \bibinfo {author} {\bibfnamefont {F.}~\bibnamefont {Meier}}, \bibinfo
  {author} {\bibfnamefont {N.}~\bibnamefont {Nurgalieva}}, \bibinfo {author}
  {\bibfnamefont {P.~P.}\ \bibnamefont {Potts}}, \bibinfo {author}
  {\bibfnamefont {R.}~\bibnamefont {Silva}},\ and\ \bibinfo {author}
  {\bibfnamefont {M.~T.}\ \bibnamefont {Mitchison}},\ }\bibfield  {title}
  {\bibinfo {title} {Optimal time estimation and the clock uncertainty relation
  for stochastic processes},\ }\href@noop {} {\bibfield  {journal} {\bibinfo
  {journal} {Physical Review X}\ }\textbf {\bibinfo {volume} {15}},\ \bibinfo
  {pages} {031068} (\bibinfo {year} {2025}{\natexlab{b}})}\BibitemShut
  {NoStop}%
\bibitem [{\citenamefont {Maes}(2020)}]{maes2020frenesy}%
  \BibitemOpen
  \bibfield  {author} {\bibinfo {author} {\bibfnamefont {C.}~\bibnamefont
  {Maes}},\ }\bibfield  {title} {\bibinfo {title} {Frenesy: Time-symmetric
  dynamical activity in nonequilibria},\ }\href@noop {} {\bibfield  {journal}
  {\bibinfo  {journal} {Physics Reports}\ }\textbf {\bibinfo {volume} {850}},\
  \bibinfo {pages} {1} (\bibinfo {year} {2020})}\BibitemShut {NoStop}%
\bibitem [{\citenamefont {Lu}\ \emph {et~al.}(1998)\citenamefont {Lu},
  \citenamefont {Xun},\ and\ \citenamefont {Xie}}]{lu1998single}%
  \BibitemOpen
  \bibfield  {author} {\bibinfo {author} {\bibfnamefont {H.~P.}\ \bibnamefont
  {Lu}}, \bibinfo {author} {\bibfnamefont {L.}~\bibnamefont {Xun}},\ and\
  \bibinfo {author} {\bibfnamefont {X.~S.}\ \bibnamefont {Xie}},\ }\bibfield
  {title} {\bibinfo {title} {Single-molecule enzymatic dynamics},\ }\href@noop
  {} {\bibfield  {journal} {\bibinfo  {journal} {Science}\ }\textbf {\bibinfo
  {volume} {282}},\ \bibinfo {pages} {1877} (\bibinfo {year}
  {1998})}\BibitemShut {NoStop}%
\bibitem [{\citenamefont {English}\ \emph {et~al.}(2006)\citenamefont
  {English}, \citenamefont {Min}, \citenamefont {Van~Oijen}, \citenamefont
  {Lee}, \citenamefont {Luo}, \citenamefont {Sun}, \citenamefont {Cherayil},
  \citenamefont {Kou},\ and\ \citenamefont {Xie}}]{english2006ever}%
  \BibitemOpen
  \bibfield  {author} {\bibinfo {author} {\bibfnamefont {B.~P.}\ \bibnamefont
  {English}}, \bibinfo {author} {\bibfnamefont {W.}~\bibnamefont {Min}},
  \bibinfo {author} {\bibfnamefont {A.~M.}\ \bibnamefont {Van~Oijen}}, \bibinfo
  {author} {\bibfnamefont {K.~T.}\ \bibnamefont {Lee}}, \bibinfo {author}
  {\bibfnamefont {G.}~\bibnamefont {Luo}}, \bibinfo {author} {\bibfnamefont
  {H.}~\bibnamefont {Sun}}, \bibinfo {author} {\bibfnamefont {B.~J.}\
  \bibnamefont {Cherayil}}, \bibinfo {author} {\bibfnamefont {S.}~\bibnamefont
  {Kou}},\ and\ \bibinfo {author} {\bibfnamefont {X.~S.}\ \bibnamefont {Xie}},\
  }\bibfield  {title} {\bibinfo {title} {Ever-fluctuating single enzyme
  molecules: Michaelis-menten equation revisited},\ }\href@noop {} {\bibfield
  {journal} {\bibinfo  {journal} {Nature chemical biology}\ }\textbf {\bibinfo
  {volume} {2}},\ \bibinfo {pages} {87} (\bibinfo {year} {2006})}\BibitemShut
  {NoStop}%
\bibitem [{\citenamefont {Vollmar}\ \emph {et~al.}(2024)\citenamefont
  {Vollmar}, \citenamefont {Bebon}, \citenamefont {Schimpf}, \citenamefont
  {Flietel}, \citenamefont {Celiksoy}, \citenamefont {S{\"o}nnichsen},
  \citenamefont {Godec},\ and\ \citenamefont {Hugel}}]{vollmar2024model}%
  \BibitemOpen
  \bibfield  {author} {\bibinfo {author} {\bibfnamefont {L.}~\bibnamefont
  {Vollmar}}, \bibinfo {author} {\bibfnamefont {R.}~\bibnamefont {Bebon}},
  \bibinfo {author} {\bibfnamefont {J.}~\bibnamefont {Schimpf}}, \bibinfo
  {author} {\bibfnamefont {B.}~\bibnamefont {Flietel}}, \bibinfo {author}
  {\bibfnamefont {S.}~\bibnamefont {Celiksoy}}, \bibinfo {author}
  {\bibfnamefont {C.}~\bibnamefont {S{\"o}nnichsen}}, \bibinfo {author}
  {\bibfnamefont {A.}~\bibnamefont {Godec}},\ and\ \bibinfo {author}
  {\bibfnamefont {T.}~\bibnamefont {Hugel}},\ }\bibfield  {title} {\bibinfo
  {title} {Model-free inference of memory in conformational dynamics of a
  multi-domain protein},\ }\href@noop {} {\bibfield  {journal} {\bibinfo
  {journal} {Journal of Physics A: Mathematical and Theoretical}\ }\textbf
  {\bibinfo {volume} {57}},\ \bibinfo {pages} {365001} (\bibinfo {year}
  {2024})}\BibitemShut {NoStop}%
\bibitem [{\citenamefont {Rothberg}\ \emph {et~al.}(1997)\citenamefont
  {Rothberg}, \citenamefont {Bello},\ and\ \citenamefont
  {Magleby}}]{rothberg1997two}%
  \BibitemOpen
  \bibfield  {author} {\bibinfo {author} {\bibfnamefont {B.~S.}\ \bibnamefont
  {Rothberg}}, \bibinfo {author} {\bibfnamefont {R.~A.}\ \bibnamefont
  {Bello}},\ and\ \bibinfo {author} {\bibfnamefont {K.~L.}\ \bibnamefont
  {Magleby}},\ }\bibfield  {title} {\bibinfo {title} {Two-dimensional
  components and hidden dependencies provide insight into ion channel gating
  mechanisms},\ }\href@noop {} {\bibfield  {journal} {\bibinfo  {journal}
  {Biophysical journal}\ }\textbf {\bibinfo {volume} {72}},\ \bibinfo {pages}
  {2524} (\bibinfo {year} {1997})}\BibitemShut {NoStop}%
\bibitem [{\citenamefont {Truccolo}\ \emph {et~al.}(2005)\citenamefont
  {Truccolo}, \citenamefont {Eden}, \citenamefont {Fellows}, \citenamefont
  {Donoghue},\ and\ \citenamefont {Brown}}]{truccolo2005point}%
  \BibitemOpen
  \bibfield  {author} {\bibinfo {author} {\bibfnamefont {W.}~\bibnamefont
  {Truccolo}}, \bibinfo {author} {\bibfnamefont {U.~T.}\ \bibnamefont {Eden}},
  \bibinfo {author} {\bibfnamefont {M.~R.}\ \bibnamefont {Fellows}}, \bibinfo
  {author} {\bibfnamefont {J.~P.}\ \bibnamefont {Donoghue}},\ and\ \bibinfo
  {author} {\bibfnamefont {E.~N.}\ \bibnamefont {Brown}},\ }\bibfield  {title}
  {\bibinfo {title} {A point process framework for relating neural spiking
  activity to spiking history, neural ensemble, and extrinsic covariate
  effects},\ }\href@noop {} {\bibfield  {journal} {\bibinfo  {journal} {Journal
  of neurophysiology}\ }\textbf {\bibinfo {volume} {93}},\ \bibinfo {pages}
  {1074} (\bibinfo {year} {2005})}\BibitemShut {NoStop}%
\bibitem [{\citenamefont {Farkhooi}\ \emph {et~al.}(2009)\citenamefont
  {Farkhooi}, \citenamefont {Strube-Bloss},\ and\ \citenamefont
  {Nawrot}}]{farkhooi2009serial}%
  \BibitemOpen
  \bibfield  {author} {\bibinfo {author} {\bibfnamefont {F.}~\bibnamefont
  {Farkhooi}}, \bibinfo {author} {\bibfnamefont {M.~F.}\ \bibnamefont
  {Strube-Bloss}},\ and\ \bibinfo {author} {\bibfnamefont {M.~P.}\ \bibnamefont
  {Nawrot}},\ }\bibfield  {title} {\bibinfo {title} {Serial correlation in
  neural spike trains: Experimental evidence, stochastic modeling, and single
  neuron variability},\ }\href@noop {} {\bibfield  {journal} {\bibinfo
  {journal} {Physical Review E—Statistical, Nonlinear, and Soft Matter
  Physics}\ }\textbf {\bibinfo {volume} {79}},\ \bibinfo {pages} {021905}
  (\bibinfo {year} {2009})}\BibitemShut {NoStop}%
\bibitem [{\citenamefont {Hummer}\ and\ \citenamefont
  {Szabo}(2015)}]{hummer2015optimal}%
  \BibitemOpen
  \bibfield  {author} {\bibinfo {author} {\bibfnamefont {G.}~\bibnamefont
  {Hummer}}\ and\ \bibinfo {author} {\bibfnamefont {A.}~\bibnamefont {Szabo}},\
  }\bibfield  {title} {\bibinfo {title} {Optimal dimensionality reduction of
  multistate kinetic and markov-state models},\ }\href@noop {} {\bibfield
  {journal} {\bibinfo  {journal} {The Journal of Physical Chemistry B}\
  }\textbf {\bibinfo {volume} {119}},\ \bibinfo {pages} {9029} (\bibinfo {year}
  {2015})}\BibitemShut {NoStop}%
\bibitem [{\citenamefont {Hartich}\ and\ \citenamefont
  {Godec}(2021)}]{hartich2021emergent}%
  \BibitemOpen
  \bibfield  {author} {\bibinfo {author} {\bibfnamefont {D.}~\bibnamefont
  {Hartich}}\ and\ \bibinfo {author} {\bibfnamefont {A.}~\bibnamefont
  {Godec}},\ }\bibfield  {title} {\bibinfo {title} {Emergent memory and kinetic
  hysteresis in strongly driven networks},\ }\href@noop {} {\bibfield
  {journal} {\bibinfo  {journal} {Physical Review X}\ }\textbf {\bibinfo
  {volume} {11}},\ \bibinfo {pages} {041047} (\bibinfo {year}
  {2021})}\BibitemShut {NoStop}%
\bibitem [{\citenamefont {Zhao}\ \emph {et~al.}(2024)\citenamefont {Zhao},
  \citenamefont {Hartich},\ and\ \citenamefont {Godec}}]{zhao2024emergence}%
  \BibitemOpen
  \bibfield  {author} {\bibinfo {author} {\bibfnamefont {X.}~\bibnamefont
  {Zhao}}, \bibinfo {author} {\bibfnamefont {D.}~\bibnamefont {Hartich}},\ and\
  \bibinfo {author} {\bibfnamefont {A.}~\bibnamefont {Godec}},\ }\bibfield
  {title} {\bibinfo {title} {Emergence of memory in equilibrium versus
  nonequilibrium systems},\ }\href@noop {} {\bibfield  {journal} {\bibinfo
  {journal} {Physical Review Letters}\ }\textbf {\bibinfo {volume} {132}},\
  \bibinfo {pages} {147101} (\bibinfo {year} {2024})}\BibitemShut {NoStop}%
\bibitem [{\citenamefont {Zhao}\ \emph {et~al.}(2025)\citenamefont {Zhao},
  \citenamefont {Makarov},\ and\ \citenamefont {Godec}}]{zhao2025towards}%
  \BibitemOpen
  \bibfield  {author} {\bibinfo {author} {\bibfnamefont {X.}~\bibnamefont
  {Zhao}}, \bibinfo {author} {\bibfnamefont {D.~E.}\ \bibnamefont {Makarov}},\
  and\ \bibinfo {author} {\bibfnamefont {A.}~\bibnamefont {Godec}},\ }\bibfield
   {title} {\bibinfo {title} {Towards markov-state holography},\ }\href@noop {}
  {\bibfield  {journal} {\bibinfo  {journal} {New Journal of Physics}\ }\textbf
  {\bibinfo {volume} {27}},\ \bibinfo {pages} {125002} (\bibinfo {year}
  {2025})}\BibitemShut {NoStop}%
\bibitem [{\citenamefont {Blom}\ \emph {et~al.}(2024)\citenamefont {Blom},
  \citenamefont {Song}, \citenamefont {Vouga}, \citenamefont {Godec},\ and\
  \citenamefont {Makarov}}]{blom2024milestoning}%
  \BibitemOpen
  \bibfield  {author} {\bibinfo {author} {\bibfnamefont {K.}~\bibnamefont
  {Blom}}, \bibinfo {author} {\bibfnamefont {K.}~\bibnamefont {Song}}, \bibinfo
  {author} {\bibfnamefont {E.}~\bibnamefont {Vouga}}, \bibinfo {author}
  {\bibfnamefont {A.}~\bibnamefont {Godec}},\ and\ \bibinfo {author}
  {\bibfnamefont {D.~E.}\ \bibnamefont {Makarov}},\ }\bibfield  {title}
  {\bibinfo {title} {Milestoning estimators of dissipation in systems observed
  at a coarse resolution},\ }\href@noop {} {\bibfield  {journal} {\bibinfo
  {journal} {Proceedings of the National Academy of Sciences}\ }\textbf
  {\bibinfo {volume} {121}},\ \bibinfo {pages} {e2318333121} (\bibinfo {year}
  {2024})}\BibitemShut {NoStop}%
\bibitem [{\citenamefont {Schwarz}\ \emph {et~al.}(2026)\citenamefont
  {Schwarz}, \citenamefont {Kolomeisky},\ and\ \citenamefont
  {Godec}}]{schwarz2026consistent}%
  \BibitemOpen
  \bibfield  {author} {\bibinfo {author} {\bibfnamefont {T.}~\bibnamefont
  {Schwarz}}, \bibinfo {author} {\bibfnamefont {A.~B.}\ \bibnamefont
  {Kolomeisky}},\ and\ \bibinfo {author} {\bibfnamefont {A.}~\bibnamefont
  {Godec}},\ }\bibfield  {title} {\bibinfo {title} {Consistent time reversal
  and reliable and accurate inference in the presence of memory},\ }\href@noop
  {} {\bibfield  {journal} {\bibinfo  {journal} {Physical Review Letters}\
  }\textbf {\bibinfo {volume} {137}},\ \bibinfo {pages} {187401} (\bibinfo
  {year} {2026})}\BibitemShut {NoStop}%
\bibitem [{\citenamefont {Vu}\ and\ \citenamefont
  {Hasegawa}(2020)}]{vu2020generalized}%
  \BibitemOpen
  \bibfield  {author} {\bibinfo {author} {\bibfnamefont {T.~V.}\ \bibnamefont
  {Vu}}\ and\ \bibinfo {author} {\bibfnamefont {Y.}~\bibnamefont {Hasegawa}},\
  }\bibfield  {title} {\bibinfo {title} {Generalized uncertainty relations for
  semi-markov processes},\ }in\ \href@noop {} {\emph {\bibinfo {booktitle}
  {Journal of Physics: Conference Series}}},\ Vol.\ \bibinfo {volume} {1593}\
  (\bibinfo {organization} {IOP Publishing},\ \bibinfo {year} {2020})\ p.\
  \bibinfo {pages} {012006}\BibitemShut {NoStop}%
\bibitem [{\citenamefont {Coghi}\ \emph {et~al.}(2026)\citenamefont {Coghi},
  \citenamefont {Budhiraja},\ and\ \citenamefont
  {Garrahan}}]{coghi2026levelprl}%
  \BibitemOpen
  \bibfield  {author} {\bibinfo {author} {\bibfnamefont {F.}~\bibnamefont
  {Coghi}}, \bibinfo {author} {\bibfnamefont {A.}~\bibnamefont {Budhiraja}},\
  and\ \bibinfo {author} {\bibfnamefont {J.~P.}\ \bibnamefont {Garrahan}},\
  }\bibfield  {title} {\bibinfo {title} {Level 2.5 large deviations and
  uncertainty relations for non-markov self-interacting dynamics},\ }\href@noop
  {} {\bibfield  {journal} {\bibinfo  {journal} {Physical Review Letters}\
  }\textbf {\bibinfo {volume} {137}},\ \bibinfo {pages} {197101} (\bibinfo
  {year} {2026})}\BibitemShut {NoStop}%
\bibitem [{\citenamefont {Coghi}\ and\ \citenamefont
  {Garrahan}(2026)}]{coghi2026level}%
  \BibitemOpen
  \bibfield  {author} {\bibinfo {author} {\bibfnamefont {F.}~\bibnamefont
  {Coghi}}\ and\ \bibinfo {author} {\bibfnamefont {J.~P.}\ \bibnamefont
  {Garrahan}},\ }\bibfield  {title} {\bibinfo {title} {Level-2.5 large
  deviations and uncertainty relations for self-interacting jump processes:
  Tilting constructions and the emergence of timescale separation},\
  }\href@noop {} {\bibfield  {journal} {\bibinfo  {journal} {Physical Review
  E}\ }\textbf {\bibinfo {volume} {114}},\ \bibinfo {pages} {054112} (\bibinfo
  {year} {2026})}\BibitemShut {NoStop}%
\bibitem [{\citenamefont {Zheng}\ and\ \citenamefont
  {Lu}(2026{\natexlab{a}})}]{zheng2026event}%
  \BibitemOpen
  \bibfield  {author} {\bibinfo {author} {\bibfnamefont {J.}~\bibnamefont
  {Zheng}}\ and\ \bibinfo {author} {\bibfnamefont {Z.}~\bibnamefont {Lu}},\
  }\bibfield  {title} {\bibinfo {title} {Event order as a resource for time
  estimation in experimental non-markovian dynamics}} (\bibinfo {year}
  {2026}{\natexlab{a}}),\ \bibinfo {note} {unpublished}\BibitemShut {NoStop}%
\bibitem [{\citenamefont {Meyer}(1962)}]{meyer1962decomposition}%
  \BibitemOpen
  \bibfield  {author} {\bibinfo {author} {\bibfnamefont {P.-A.}\ \bibnamefont
  {Meyer}},\ }\bibfield  {title} {\bibinfo {title} {A decomposition theorem for
  supermartingales},\ }\href@noop {} {\bibfield  {journal} {\bibinfo  {journal}
  {Illinois Journal of Mathematics}\ }\textbf {\bibinfo {volume} {6}},\
  \bibinfo {pages} {193} (\bibinfo {year} {1962})}\BibitemShut {NoStop}%
\bibitem [{\citenamefont {Yang}\ and\ \citenamefont
  {Qian}(2023)}]{yang2023time}%
  \BibitemOpen
  \bibfield  {author} {\bibinfo {author} {\bibfnamefont {Y.-J.}\ \bibnamefont
  {Yang}}\ and\ \bibinfo {author} {\bibfnamefont {H.}~\bibnamefont {Qian}},\
  }\bibfield  {title} {\bibinfo {title} {Time-translational symmetry in
  statistical dynamics dictates einstein relation, green-kubo formula, and
  their generalizations},\ }\href@noop {} {\bibfield  {journal} {\bibinfo
  {journal} {Physical Review E}\ }\textbf {\bibinfo {volume} {107}},\ \bibinfo
  {pages} {024110} (\bibinfo {year} {2023})}\BibitemShut {NoStop}%
\bibitem [{\citenamefont {Stutzer}\ \emph {et~al.}(2026)\citenamefont
  {Stutzer}, \citenamefont {Dieball},\ and\ \citenamefont
  {Godec}}]{stutzer2026stochastic}%
  \BibitemOpen
  \bibfield  {author} {\bibinfo {author} {\bibfnamefont {L.~T.}\ \bibnamefont
  {Stutzer}}, \bibinfo {author} {\bibfnamefont {C.}~\bibnamefont {Dieball}},\
  and\ \bibinfo {author} {\bibfnamefont {A.}~\bibnamefont {Godec}},\ }\bibfield
   {title} {\bibinfo {title} {Stochastic calculus for pathwise observables of
  markov-jump processes: Unification of diffusion and jump dynamics},\
  }\href@noop {} {\bibfield  {journal} {\bibinfo  {journal} {Physical Review
  X}\ }\textbf {\bibinfo {volume} {16}},\ \bibinfo {pages} {021038} (\bibinfo
  {year} {2026})}\BibitemShut {NoStop}%
\bibitem [{\citenamefont {Zheng}\ and\ \citenamefont
  {Lu}(2026{\natexlab{b}})}]{zheng2026mutual}%
  \BibitemOpen
  \bibfield  {author} {\bibinfo {author} {\bibfnamefont {J.}~\bibnamefont
  {Zheng}}\ and\ \bibinfo {author} {\bibfnamefont {Z.}~\bibnamefont {Lu}},\
  }\bibfield  {title} {\bibinfo {title} {Mutual linearity in and out of
  stationarity for markov jump processes: A trajectory-based approach},\
  }\href@noop {} {\bibfield  {journal} {\bibinfo  {journal} {Physical Review
  E}\ }\textbf {\bibinfo {volume} {114}},\ \bibinfo {pages} {014126} (\bibinfo
  {year} {2026}{\natexlab{b}})}\BibitemShut {NoStop}%
\bibitem [{\citenamefont {Zheng}\ and\ \citenamefont
  {Lu}(2026{\natexlab{c}})}]{zheng2026memory}%
  \BibitemOpen
  \bibfield  {author} {\bibinfo {author} {\bibfnamefont {J.}~\bibnamefont
  {Zheng}}\ and\ \bibinfo {author} {\bibfnamefont {Z.}~\bibnamefont {Lu}},\
  }\bibfield  {title} {\bibinfo {title} {The memory hidden in response
  fluctuations: Trajectory-level fluctuation-response theory and inequalities
  for non-markovian jump dynamics},\ }\href@noop {} {\bibfield  {journal}
  {\bibinfo  {journal} {arXiv preprint arXiv:2608.20328}\ } (\bibinfo {year}
  {2026}{\natexlab{c}})}\BibitemShut {NoStop}%
\bibitem [{\citenamefont {Le~Gall}(2016)}]{le2016brownian}%
  \BibitemOpen
  \bibfield  {author} {\bibinfo {author} {\bibfnamefont {J.-F.}\ \bibnamefont
  {Le~Gall}},\ }\href@noop {} {\emph {\bibinfo {title} {Brownian motion,
  martingales, and stochastic calculus}}},\ Vol.\ \bibinfo {volume} {274}\
  (\bibinfo  {publisher} {Springer},\ \bibinfo {year} {2016})\BibitemShut
  {NoStop}%
\bibitem [{sup()}]{supp}%
  \BibitemOpen
  \href@noop {} {}\bibinfo {note} {See Supplemental Material at [URL] for more
  details.}\BibitemShut {Stop}%
\bibitem [{\citenamefont {Zwanzig}(1990)}]{zwanzig1990rate}%
  \BibitemOpen
  \bibfield  {author} {\bibinfo {author} {\bibfnamefont {R.}~\bibnamefont
  {Zwanzig}},\ }\bibfield  {title} {\bibinfo {title} {Rate processes with
  dynamical disorder},\ }\href@noop {} {\bibfield  {journal} {\bibinfo
  {journal} {Accounts of Chemical Research}\ }\textbf {\bibinfo {volume}
  {23}},\ \bibinfo {pages} {148} (\bibinfo {year} {1990})}\BibitemShut
  {NoStop}%
\bibitem [{\citenamefont {Zwanzig}(1992)}]{10.1063/1.462993}%
  \BibitemOpen
  \bibfield  {author} {\bibinfo {author} {\bibfnamefont {R.}~\bibnamefont
  {Zwanzig}},\ }\bibfield  {title} {\bibinfo {title} {Dynamical disorder:
  Passage through a fluctuating bottleneck},\ }\href
  {https://doi.org/10.1063/1.462993} {\bibfield  {journal} {\bibinfo  {journal}
  {The Journal of Chemical Physics}\ }\textbf {\bibinfo {volume} {97}},\
  \bibinfo {pages} {3587} (\bibinfo {year} {1992})}\BibitemShut {NoStop}%
\bibitem [{\citenamefont {Zheng}(2026)}]{data}%
  \BibitemOpen
  \bibfield  {author} {\bibinfo {author} {\bibfnamefont {J.}~\bibnamefont
  {Zheng}},\ }\href@noop {} {} (\bibinfo {year} {2026}),\ \bibinfo {note}
  {https://github.com/Axeho2/Kinetic-and-Clock-Uncertainty-Relations-in-General-Non-Markovian-Jump-Dynamics.}\BibitemShut
  {Stop}%
\end{thebibliography}%

\end{document}


\preprint{APS/123-QED}

\title{Supplemental Material for ``Kinetic and Clock Uncertainty Relations in General Non-Markovian Jump Dynamics''}
\author{Jiming Zheng} \email{jiming@unc.edu}
\author{Zhiyue Lu}
\affiliation{Department of Chemistry, University of North Carolina-Chapel Hill, NC}

\date{\today}

\maketitle

In this Supplemental Material, we provide detailed derivations supporting the results in the main text. In \cref{SMsec: trajectory_probability}, we derive the trajectory score under complete time dilation using the martingale structure of non-Markovian jump processes. In \cref{SMsec: finite_time_uncertainty}, we establish the finite-time uncertainty relations and discuss their optimality. In \cref{SMsec: illustration}, we analyze the eight-state dynamic-disorder model. Specifically, \cref{SMsubsec: disorder_counting} derives the counting precision bound, while \cref{SMsubsec: disorder_order} establishes the additional timing information carried by event order and its exponential advantage. Finally, \cref{SMsec: driven_time_dilation,SMsec: timestamped_observables} extend the results to externally driven dynamics and time-tagged observables, respectively.

\tableofcontents

\newpage

\section{Trajectory probability and complete time dilation}
\label{SMsec: trajectory_probability}

Consider the observed trajectory $X_\tau=(x_0;\{e_k,t_k\}_{k=1}^{n_\tau})$ with $0<t_1<\cdots<t_{n_\tau}<\tau$.
The conditional jump intensity of channel $e$ is $\gamma_e^{X_{t-}}(t)$, and its compensated counting increment is
\begin{equation}
  \mathrm{d}\varepsilon_e(t)
  = \mathrm{d}n_e(t) - \gamma_e^{X_{t-}}(t)\mathrm{d}t.
  \label{SMeq: compensated_counting_increment}
\end{equation}
We choose the initial preparation to be compatible with complete time dilation and to have a $c$-independent probability law.
For histories extending before $t=0$, the initial prehistory must be stretched together with the observed trajectory; if its probability law depends on $c$, the logarithmic derivative of that law contributes an additional initial score.
The formulas below apply directly when this initial score vanishes, as for a fixed preparation with memory initialized at $t=0$ or a scale-covariant microscopic initial law.
We also assume that the intensities are positive on allowed transitions, differentiable under time dilation, and sufficiently regular for the following differentiation and martingale isometry.

For a fixed observed trajectory, the complete time-dilated intensity is
\begin{equation}
  \gamma_{e,c}^{X_{t-}}(t)
  = c\gamma_e^{\mathfrak{T}_c X_{t-}}(ct),
  \qquad
  (\mathfrak{T}_c X_{t-})(s) = x_{s/c},\quad 0\leq s<ct.
  \label{SMeq: complete_time_dilation}
\end{equation}
The factor $c$ rescales the physical rate of jumps, while the transformation of the history rescales every internal memory time scale.
The likelihood of $X_\tau$, relative to the usual product measure of jump times and channels, is
\begin{equation}
  \mathcal{P}_c^X[X_\tau]
  = p_0(x_0)
  \prod_{k=1}^{n_\tau}\gamma_{e_k,c}^{X_{t_k-}}(t_k)
  \exp\left[-\sum_e\int_0^\tau\gamma_{e,c}^{X_{t-}}(t)\mathrm{d}t\right],
  \label{SMeq: dilated_path_probability}
\end{equation}
where the initial factor is understood to include any fixed initial memory condition.
Taking the logarithmic derivative of \cref{SMeq: dilated_path_probability} gives
\begin{equation}
  \begin{aligned}
  \left.\partial_{\ln c}\ln \mathcal{P}_c^X[X_\tau]\right|_{c=1}
  ={}& \sum_e\int_0^\tau
  \left.\partial_{\ln c}\ln\gamma_{e,c}^{X_{t-}}(t)\right|_{c=1}
  \mathrm{d}n_e(t) \\
  &-\sum_e\int_0^\tau
  \left.\partial_{\ln c}\gamma_{e,c}^{X_{t-}}(t)\right|_{c=1}
  \mathrm{d}t.
  \end{aligned}
  \label{SMeq: differentiated_path_probability}
\end{equation}
Using the definition of $\mu_e(t)$ in the main text,
\begin{equation}
  \left.\partial_{\ln c}\ln\gamma_{e,c}^{X_{t-}}(t)\right|_{c=1}
  = 1+\mu_e(t),
  \label{SMeq: dilation_sensitivity}
\end{equation}
and $\partial_{\ln c}\gamma_{e,c}|_{c=1}=\gamma_e^{X_{t-}}(1+\mu_e)$, we obtain
\begin{equation}
  \begin{aligned}
  \Lambda_\tau^X
  &= \sum_e\int_0^\tau[1+\mu_e(t)]
     \left[\mathrm{d}n_e(t)-\gamma_e^{X_{t-}}(t)\mathrm{d}t\right] \\
  &= \sum_e\int_0^\tau[1+\mu_e(t)]\mathrm{d}\varepsilon_e(t).
  \end{aligned}
  \label{SMeq: complete_score}
\end{equation}
The coefficient $1+\mu_e(t)$ is predictable because it is determined by the history preceding the jump. Consequently, $\Lambda_\tau^X$ has zero mean. Since $1+\mu_e(t)$ is predictable, the martingale representation gives $\langle \Lambda_\tau^X \rangle = 0$. The martingale isometry further yields
\begin{equation}
  \mathcal{I}_\tau^X
  = \left\langle (\Lambda_\tau^X)^2 \right\rangle
  = \sum_e \left\langle
  \int_0^\tau \gamma_e(t)[1+\mu_e(t)]^2 \rmd t
  \right\rangle,
\end{equation}
which gives the local expression for the full-record Fisher information in the main text.

As an example, consider an allowed transition with
$\gamma_e^{X_{t-}}(t)=r_e(1+a_t/\tau_m)$, where $a_t$ is the age since the last jump.
Time dilation replaces $a_t$ by $ca_t$ inside the intensity, and therefore
\begin{equation}
  \mu_e(t)
  =\left.\partial_{\ln c}\ln(1+ca_t/\tau_m)\right|_{c=1}
  =\frac{a_t}{\tau_m+a_t}.
  \label{SMeq: age_dependent_sensitivity}
\end{equation}
No particular form of the history dependence is needed for \cref{SMeq: complete_score}.

\section{Finite-time uncertainty relations and local optimality}
\label{SMsec: finite_time_uncertainty}

Since $\mathcal{O}_\tau$ is obtained from $X_\tau$ by discarding information, differentiating the marginal likelihood gives
\begin{equation}
  \Lambda_\tau^\mathcal{O}
  = \mathbb{E}[\Lambda_\tau^X \,|\, \mathcal{O}_\tau].
\end{equation}
For two nested records $\mathcal{O}_{1,\tau}$ and
$\mathcal{O}_{2,\tau}$, where the former is determined by the latter, the law of total variance gives
\begin{equation}
  \mathcal{I}_\tau^{\mathcal{O}_2}
  - \mathcal{I}_\tau^{\mathcal{O}_1}
  = \mathbb{E}\!\left[
    \mathrm{Var}\!\left(
      \Lambda_\tau^{\mathcal{O}_2}
      \,|\, \mathcal{O}_{1,\tau}
    \right)
  \right].
\end{equation}

We first consider an autonomous process and a record observable $Q_\tau^O=Q(O_\tau)$ that does not explicitly depend on the observation horizon or on the waiting times.
Complete time dilation maps a trajectory observed up to $\tau$ to an undilated trajectory observed up to $c\tau$, without changing the value of this observable on the corresponding sequence of jumps.
Therefore
\begin{equation}
  \langle Q_\tau^O\rangle_c
  =\langle Q_{c\tau}^O\rangle,
  \qquad
  \left.\partial_{\ln c}\langle Q_\tau^O\rangle_c\right|_{c=1}
  =\tau\partial_\tau\langle Q_\tau^O\rangle.
  \label{SMeq: time_scaling_mean_identity}
\end{equation}
On the other hand, differentiating the expectation using the likelihood of the observed record gives
\begin{equation}
  \begin{aligned}
  \left.\partial_{\ln c}\langle Q_\tau^O\rangle_c\right|_{c=1}
  &=\langle Q_\tau^O\Lambda_\tau^O\rangle\\
  &=\langle(Q_\tau^O-\langle Q_\tau^O\rangle)\Lambda_\tau^O\rangle.
  \end{aligned}
  \label{SMeq: score_response_identity}
\end{equation}
The last equality follows from the zero mean of the score.
Combining \cref{SMeq: time_scaling_mean_identity} and \eqref{SMeq: score_response_identity} yields the identity in Eq.~(10) of the main text.
The Cauchy-Schwarz inequality gives
\begin{equation}
  [ \tau \partial_\tau \langle Q_\tau^O \rangle]^2 \leq \Var[Q_\tau^O] \mathcal{I}_\tau^O,
  \label{SMeq: finite_time_bound}
\end{equation}
which includes the non-Markovian KUR, CUR, and counting inequality when the corresponding record is chosen.

Equality in the score-variance Cauchy-Schwarz inequality holds when
\begin{equation}
  Q_\tau^O - \langle Q_\tau^O \rangle = C \Lambda_\tau^O
  \label{SMeq: local_saturation}
\end{equation}
for a nonzero constant $C$.
Strictly speaking, this is a local statement at a specified observation horizon $\tau_0$: the score $\Lambda_{\tau_0}^O$ is used to define a fixed readout function of the record, calibrated at $\tau_0$.
The readout function is then held fixed when differentiating the dynamics with respect to time dilation.
This distinction avoids introducing an explicit horizon dependence into an otherwise untimed observable.

For an unbiased time estimator with $\langle Q_\tau^O\rangle=\tau$ over the estimation range, \cref{SMeq: finite_time_bound} becomes
\begin{equation}
  \Var[Q_\tau^O] \geq \frac{\tau^2}{\mathcal{I}_\tau^O}.
  \label{SMeq: finite_time_clock_error}
\end{equation}
If $\mathcal{I}_\tau^O / \tau \to \mathcal{T}_O^{-1} > 0$, it follows that
\begin{equation}
  \liminf_{\tau\to\infty} \frac{\Var[Q_\tau^O]}{\tau} \geq \mathcal{T}_O.
  \label{SMeq: asymptotic_clock_error}
\end{equation}
The same conclusion holds for estimators satisfying
\begin{equation}
  \langle Q_\tau^\mathcal{O} \rangle - \tau = o(\sqrt{\tau}), \qquad \partial_\tau \langle Q_\tau^\mathcal{O} \rangle = 1 + o(1),
\end{equation}
as $\tau \to \infty$. The first condition ensures that the squared bias is sublinear in $\tau$, so that the mean-square timing error and the variance have the same asymptotic growth rate. The second condition ensures that the variance bound in Eq.~(S13) remains valid.

\section{Illustration}
\label{SMsec: illustration}

\subsection{Stationary statistics and the counting bound}
\label{SMsubsec: disorder_counting}

We consider the eight-state model in the main text, with microscopic state
$(s_1,s_2,z)$ and observed state $x=s_1+s_2$.
The hidden gate $z\in\{-1,+1\}$ switches symmetrically at rate $\kappa$,
setting the first molecular rate to $a_-$ or $a_+$, respectively.
We initialize the system in equilibrium. Writing
\begin{subequations}
\label{SMeq: switch_equilibrium_probabilities}
\begin{align}
  p&\equiv\frac{\eta}{1+\eta},
  \qquad q\equiv\frac{1}{1+\eta},
  \\
  \pi(s_1,s_2,z)
  &=\frac12 p^{s_1+s_2}q^{2-s_1-s_2}
  =\frac{\eta^{s_1+s_2}}{2(1+\eta)^2},
  \label{SMeq: eight_state_stationary_law}
\end{align}
\end{subequations}
detailed balance holds on every microscopic edge.
The equilibrium distribution is independent of $a_\pm$, $b$, and $\kappa$.

At fixed $z$, the mean fluxes through the observed channels
$0\to1$, $1\to0$, $1\to2$, and $2\to1$ form the vector
\begin{align}
  \boldsymbol f_z
  &=(a_z+b)\frac{(\eta,\eta,\eta^2,\eta^2)^{\mathsf T}}
  {(1+\eta)^2}.
  \label{SMeq: observed_channel_flux_vector}
\end{align}
The differences $n_{01}-n_{10}$ and $n_{12}-n_{21}$ are fixed by the
boundary states. Thus the two counts
$\boldsymbol n_\tau=(n_{01}(\tau),n_{12}(\tau))^{\mathsf T}$,
together with these states, determine the complete count record.
Their mean rate is
\begin{subequations}
\label{SMeq: counting_covariance_matrix}
\begin{align}
  \boldsymbol u
  &\equiv\frac{1}{(1+\eta)^2}
  \begin{pmatrix}\eta\\ \eta^2\end{pmatrix},
  \qquad\bar a\equiv\frac{a_++a_-}{2},
  \\
  \boldsymbol j
  &\equiv\lim_{\tau\to\infty}
  \frac{\langle\boldsymbol n_\tau\rangle}{\tau}
  =(\bar a+b)\boldsymbol u.
\end{align}
\end{subequations}
Conditioned on the entire gate trajectory, the two molecular switches
remain in their common equilibrium distribution at every time. Hence
\begin{align}
  \mathbb E[\boldsymbol n_\tau\,|\,z_{[0,\tau]}]
  &=\boldsymbol u\int_0^\tau(a_{z_t}+b)\,\rmd t.
  \label{SMeq: gated_conditional_event_mean}
\end{align}
Since $\langle z_tz_0\rangle=e^{-2\kappa|t|}$,
\begin{align}
  \lim_{\tau\to\infty}\frac1\tau
  \Var\!\left[\int_0^\tau(a_{z_t}+b)\,\rmd t\right]
  &=\frac{(a_+-a_-)^2}{4\kappa}.
  \label{SMeq: gate_integrated_variance}
\end{align}
The law of total covariance therefore gives
\begin{subequations}
\label{SMeq: count_covariance_decomposition}
\begin{align}
  \boldsymbol D
  &\equiv\lim_{\tau\to\infty}
  \frac{\Cov[\boldsymbol n_\tau]}{\tau}
  =\boldsymbol D_0+
  \frac{(a_+-a_-)^2}{4\kappa}
  \boldsymbol u\boldsymbol u^{\mathsf T},
  \\
  \boldsymbol D_0
  &\equiv\lim_{\tau\to\infty}\frac1\tau
  \mathbb E[\Cov[\boldsymbol n_\tau\,|\,z_{[0,\tau]}]].
\end{align}
\end{subequations}
At fixed positive molecular rates, molecular relaxation remains finite
as $\kappa\to0$. The matrix $\boldsymbol D_0$ is positive definite
for $\kappa>0$ and remains bounded in this limit.

For each fixed $\kappa>0$, the counts are additive observables of a
finite irreducible Markov chain. The local Gaussian expansion of their
likelihood follows from the tilted generator. Under complete time
dilation, the leading mean and covariance become $c\tau\boldsymbol j$
and $c\tau\boldsymbol D$. The mean shift gives the extensive
information, while the covariance change contributes only $O(1)$.
Thus
\begin{align}
  \mathcal T_N^{-1}
  &=\boldsymbol j^{\mathsf T}\boldsymbol D^{-1}\boldsymbol j.
  \label{SMeq: count_information_inversion}
\end{align}
This is the information rate of the full count distribution, including
nonlinear readouts. Boundary states do not change this long-time rate.

Applying the Sherman-Morrison identity gives
\begin{subequations}
\label{SMeq: counting_time_scale}
\begin{align}
  \mathcal T_N
  &=\frac{(a_+-a_-)^2}{4\kappa(\bar a+b)^2}
  +\frac{1}{(\bar a+b)^2
  \boldsymbol u^{\mathsf T}\boldsymbol D_0^{-1}\boldsymbol u},
  \\
  &=\frac1\kappa
  \left(\frac{a_+-a_-}{a_++a_-+2b}\right)^2+O(1).
\end{align}
\end{subequations}
The mean count rate and the slow gate fluctuations point in the same
direction in count space. Unordered counts therefore cannot remove
the leading timing fluctuations caused by the slowly switching gate.
The leading coefficient is independent of $\eta$.

\subsection{A reference system with a fixed catalytic rate}
\label{SMsubsec: disorder_reference}

We first determine the sequence timing scale with the gate frozen at
a fixed rate parameter $a>0$.
The observed states $x=0$ and $x=2$ uniquely identify the microscopic
states $(0,0)$ and $(1,1)$. At $x=1$, the microscopic state may instead
be $(1,0)$ or $(0,1)$, with respective escape rates
\begin{equation}
  U\equiv a+\eta b,\qquad V\equiv b+\eta a.
  \label{SMeq: middle_state_escape_rates}
\end{equation}
A complete excursion has the form $u\to1\to v$, where
$u,v\in\{0,2\}$. Each endpoint resets the microscopic state.
For a fixed number of excursions, conditioning on their endpoint
sequence therefore makes their hidden paths and durations independent.

Let $J_{uv}^{10}$ and $J_{uv}^{01}$ denote the stationary rates of
excursions through $(1,0)$ and $(0,1)$, respectively. They are
\begin{equation}
  \begin{array}{c|cc}
  (u,v)&J_{uv}^{10}&J_{uv}^{01} \\
  \midrule
  (0,0)&\displaystyle\frac{q^2\eta a^2}{U}
        &\displaystyle\frac{q^2\eta b^2}{V}\\[10pt]
  (0,2)&\displaystyle\frac{q^2\eta^2 ab}{U}
        &\displaystyle\frac{q^2\eta^2 ab}{V}\\[10pt]
  (2,0)&\displaystyle\frac{p^2 ab}{U}
        &\displaystyle\frac{p^2 ab}{V}\\[10pt]
  (2,2)&\displaystyle\frac{p^2\eta b^2}{U}
        &\displaystyle\frac{p^2\eta a^2}{V}
  \end{array}
  \label{SMeq: frozen_gate_excursion_fluxes}
\end{equation}
For example, the first entry is the equilibrium rate of
$(0,0)\to(1,0)$ multiplied by the probability that the next jump
returns to $(0,0)$.

Writing $J_{uv}\equiv J_{uv}^{10}+J_{uv}^{01}$, the conditional
density of the waiting time $w$ at $x=1$ is
\begin{equation}
  \psi_{uv}(w)
  =\frac{J_{uv}^{10}Ue^{-Uw}+J_{uv}^{01}Ve^{-Vw}}{J_{uv}},
  \qquad w\geq0.
  \label{SMeq: frozen_middle_waiting_density}
\end{equation}
Its mean and variance are
\begin{subequations}
\label{SMeq: middle_waiting_conditional_variance}
\begin{align}
  \overline w_{uv}
  &\equiv\mathbb E[w\,|\,u,1,v]
  =\frac{1}{J_{uv}}
    \left(\frac{J_{uv}^{10}}{U}+\frac{J_{uv}^{01}}{V}\right),\\
  V_{uv}
  &\equiv\Var[w\,|\,u,1,v]
  =\frac{2}{J_{uv}}
    \left(\frac{J_{uv}^{10}}{U^2}+\frac{J_{uv}^{01}}{V^2}\right)
    -\overline w_{uv}^{\,2}.
\end{align}
\end{subequations}
Let $C_{uv}(\tau)$ count completed excursions $u\to1\to v$.
The full excursion duration $H$ includes the waiting time at the
first endpoint. With $\lambda_0=\eta(a+b)$ and $\lambda_2=a+b$,
\begin{subequations}
\label{SMeq: excursion_duration_moments}
\begin{align}
  d_{uv}
  &\equiv\mathbb E[H\,|\,u,v]
  =\lambda_u^{-1}+\overline w_{uv},
  \\
  v_{uv}
  &\equiv\Var[H\,|\,u,v]
  =\lambda_u^{-2}+V_{uv}.
\end{align}
\end{subequations}
The likelihood of an untimed record is a convolution of the joint
excursion-duration densities and the boundary factors.
Admissible endpoint sequences with the same excursion counts and
boundary labels therefore have the same likelihood.
The four excursion counts, together with the labels of the initial
and final partial excursions, retain all the timing information
in the sequence.

Write
$\boldsymbol C_\tau=(C_{00}(\tau),C_{02}(\tau),C_{20}(\tau),C_{22}(\tau))^{\mathsf T}$,
and collect $J_{uv}$ and $d_{uv}$ into vectors $\boldsymbol J$ and
$\boldsymbol d$ in the same order. Define
\begin{subequations}
\label{SMeq: excursion_covariance}
\begin{align}
  \boldsymbol D^{\rm exc}
  &\equiv\lim_{\tau\to\infty}
  \frac{\Cov[\boldsymbol C_\tau]}{\tau},
  \\
  \mathcal V(a)&\equiv\sum_{u,v}J_{uv}v_{uv}.
\end{align}
\end{subequations}
Including the last visited endpoint in the microscopic state makes
the excursion counts additive observables of a finite-state Markov
chain. The same calculation as in
\cref{SMeq: count_information_inversion} gives
$\bigl(\mathcal T_\Sigma^{(a)}\bigr)^{-1}
=\boldsymbol J^{\mathsf T}(\boldsymbol D^{\rm exc})^+\boldsymbol J$.
Here $+$ denotes the Moore-Penrose pseudoinverse;
$C_{02}-C_{20}$ is fixed by the endpoint labels.

To evaluate this expression, write the duration of excursion $k$ as
$H_k=d_{u_kv_k}+\xi_k$.
For a fixed number of excursions, the residuals $\xi_k$ are independent
and centered conditional on the endpoint sequence.
Since $\boldsymbol J^{\mathsf T}\boldsymbol d=1$, weighting the
centered count reward $\boldsymbol e_{u_kv_k}-\boldsymbol JH_k$
by $\boldsymbol d$ gives $-\xi_k$, where $\boldsymbol e_{uv}$
selects excursion type $(u,v)$.
Only the residual-duration term $-\boldsymbol J\xi_k$ contributes
to its covariance with $-\xi_k$. The resulting covariance identities are
\begin{subequations}
\label{SMeq: excursion_covariance_identity}
\begin{align}
  \boldsymbol D^{\rm exc}\boldsymbol d
  &=\mathcal V(a)\boldsymbol J,
  \\
  \boldsymbol J^{\mathsf T}
  (\boldsymbol D^{\rm exc})^+\boldsymbol J
  &=\frac{1}{\mathcal V(a)}.
\end{align}
\end{subequations}
Initial and final partial excursions have bounded duration moments
and do not affect these rates. Consequently,
\begin{align}
  \mathcal T_\Sigma^{(a)}
  &=\mathcal V(a)
  =\frac{q^2}{\eta(a+b)}
  +\frac{p^2}{a+b}
  +\sum_{u,v\in\{0,2\}}J_{uv}V_{uv}.
  \label{SMeq: frozen_gate_sequence_scale}
\end{align}
An efficient clock assigns each completed excursion its conditional
mean duration:
\begin{subequations}
\label{SMeq: frozen_gate_efficient_clock}
\begin{align}
  Q_\tau^{(a)}&\equiv\sum_{u,v}d_{uv}C_{uv}(\tau),
  \\
  \langle Q_\tau^{(a)}\rangle&=\tau+O(1),
  \qquad\partial_\tau\langle Q_\tau^{(a)}\rangle\longrightarrow1,
  \\
  \lim_{\tau\to\infty}\frac{\Var[Q_\tau^{(a)}]}{\tau}
  &=\mathcal T_\Sigma^{(a)}.
\end{align}
\end{subequations}
All quantities depend explicitly on $a,b,\eta$; for example,
$\mathcal T_\Sigma^{(a)}=1/(a+b)$ when $\eta=1$.
For the two frozen gate modes, we write
\begin{align}
  \mathcal T_\Sigma^{(+)}
  &\equiv\mathcal T_\Sigma^{(a_+)},
  \qquad
  \mathcal T_\Sigma^{(-)}
  \equiv\mathcal T_\Sigma^{(a_-)}.
  \label{SMeq: frozen_gate_reference_scales}
\end{align}

\subsection{Information retained by event order}
\label{SMsubsec: disorder_order}

For a frozen rate $a$, the probability that $0\to1$ is followed
by $1\to2$ is
\begin{align}
  \theta(a)
  &\equiv\frac{J_{02}}{J_{00}+J_{02}}
  =\frac{\eta(1+\eta)ab}{(a+\eta b)(b+\eta a)}.
  \label{SMeq: gate_sequence_signature}
\end{align}
The endpoint transition probabilities are invariant under
$a\mapsto b^2/a$. We assume $a_+\ne a_-$ and $a_+a_-\ne b^2$,
so that the two gate modes can be distinguished from their
endpoint transition statistics.

First suppose the observer also knows the gate state at each visit
to $x=0$ or $x=2$.
The augmented endpoint $(u,z)$ specifies the microscopic state,
so the preceding excursion construction applies.
The corresponding efficient clock $Q_\tau^{\rm aug}$ sums the
conditional mean excursion durations given the successive
augmented endpoints.
An excursion contains a gate change with probability $O(\kappa/b)$,
and its duration moments remain bounded as $\kappa\to0$.
Excursions within mode $z$ have rates tending to $J_{uv}(a_z)/2$,
where $1/2$ is the fraction of physical time spent in that mode.
Therefore,
\begin{align}
  \mathcal T_{\rm aug}
  &=\frac{\mathcal T_\Sigma^{(+)}
        +\mathcal T_\Sigma^{(-)}}{2}
  +O(\kappa/b^2).
  \label{SMeq: augmented_gate_scale}
\end{align}
Complete time dilation scales all rates, including $\kappa$, together,
leaving the augmented endpoint transition probabilities unchanged.
Discarding the gate labels cannot increase information, so
$\mathcal T_\Sigma\geq\mathcal T_{\rm aug}$.

For a matching upper bound, divide the observed sequence into blocks
of $L_\kappa=\lceil(b/\kappa)^{1/3}\rceil$ complete excursions.
Infer the gate mode from the endpoint transition frequencies in each
block, and sum the corresponding frozen mean durations $d_{uv}(a_z)$.
Dividing by the stationary mean growth rate gives a calibrated clock
$Q_\tau^{\rm dec}$. Its readout depends only on the untimed event sequence.
At each fixed $\kappa>0$, both clocks have mean $\tau+O(1)$ and
mean derivative tending to one.

Because the two frozen endpoint chains are distinct, a block without
a gate change is misclassified with probability at most
$K_0e^{-\alpha_0L_\kappa}$.
A block contains a gate change with probability
$O(\kappa L_\kappa/b)$ and then has squared timing error
$O(L_\kappa^2/b^2)$.
Blocks occur at rate $O(b/L_\kappa)$, giving an error variance rate
$O(\kappa L_\kappa^2/b^2)$.
Correlations of the mode-dependent mean error over a gate residence
of order $\kappa^{-1}$ give the same scaling.
Misclassification errors in blocks without gate changes contribute
at most $O(e^{-\alpha_0L_\kappa}/\kappa)$, after absorbing polynomial
block factors into a smaller positive $\alpha_0$.
Together,
\begin{subequations}
\label{SMeq: decoded_gate_error}
\begin{align}
  \limsup_{\tau\to\infty}
  \frac{\langle
  (Q_\tau^{\rm dec}-Q_\tau^{\rm aug})^2\rangle}{\tau}
  &\leq\epsilon_\kappa,
  \\
  \epsilon_\kappa
  \equiv\frac{K_1}{b}\left[
  \frac{\kappa}{b}L_\kappa^2+
  \frac{b}{\kappa}e^{-\alpha_0L_\kappa}\right]
  &\longrightarrow0.
\end{align}
\end{subequations}
The positive constants $K_0,K_1,\alpha_0$ depend only on the fixed
ratios $a_\pm/b$ and $\eta$.
Calibration removes the asymptotic mean drift, and bounded boundary
corrections do not affect this rate.
The $O(\kappa/b^2)$ corrections to the frozen mean durations are
also covered by the first term.

The clock inequality and Cauchy-Schwarz give
\begin{align}
  \mathcal T_{\rm aug}
  \leq\mathcal T_\Sigma
  &\leq\limsup_{\tau\to\infty}
  \frac{\Var[Q_\tau^{\rm dec}]}{\tau}
  \leq
  \left(\sqrt{\mathcal T_{\rm aug}}
        +\sqrt{\epsilon_\kappa}\right)^2.
  \label{SMeq: slow_gate_precision_sandwich}
\end{align}
Combining this with \cref{SMeq: augmented_gate_scale} yields
\begin{align}
  \mathcal T_\Sigma
  &=\frac{\mathcal T_\Sigma^{(+)}
        +\mathcal T_\Sigma^{(-)}}{2}+o(1),
  \qquad\kappa\to0.
  \label{SMeq: slow_gate_ordered_scale}
\end{align}
Event order therefore removes the divergent contribution of slow
gate fluctuations: $\mathcal T_\Sigma$ remains finite while
$\mathcal T_N$ grows as $\kappa^{-1}$.

For thermally activated switching,
$\kappa=\kappa_0\exp[-B/(k_{\rm B}T)]$, increasing the barrier at
fixed temperature, prefactor, and molecular rates gives
\begin{align}
  \ln\frac{\mathcal T_\Sigma}{\mathcal T_N}
  &=-\frac{B}{k_{\rm B}T}+O(1),
  \qquad B/(k_{\rm B}T)\to\infty.
  \label{SMeq: exponential_sequence_advantage}
\end{align}
The long-time limit $\tau\to\infty$ is taken at each fixed
$\kappa>0$ before $\kappa\to0$.
The precision comparison therefore applies in the regime
$\tau\gg\kappa^{-1}$.

\section{Time dilation in externally driven dynamics}
\label{SMsec: driven_time_dilation}

Suppose that the jump intensities depend on an externally specified protocol $u_v(t)\equiv u(vt)$.
A complete speed-up of the internal dynamics must leave the laboratory protocol unchanged.
The auxiliary intensity is therefore
\begin{equation}
  \gamma_{e,v}^{Y_{t-}}(t;c)
  =c\gamma_{e,v/c}^{\mathfrak{T}_cY_{t-}}(ct).
  \label{SMeq: driven_auxiliary_intensity}
\end{equation}
The protocol in the auxiliary trajectory is indeed unchanged because
\begin{equation}
  u_{v/c}(ct)=u(vt)=u_v(t).
  \label{SMeq: driven_protocol_invariance}
\end{equation}
For an observable determined by the untimed event record, the auxiliary process up to laboratory time $\tau$ has the same record distribution as the original process observed up to $c\tau$ with driving speed $v/c$.
Thus
\begin{equation}
  \langle Q_\tau^O\rangle_{c,v}
  =\langle Q_{c\tau}^O\rangle_{v/c},
  \label{SMeq: driven_time_scaling_identity}
\end{equation}
and differentiation at $c=1$ yields
\begin{equation}
  \left.\partial_{\ln c}\langle Q_\tau^O\rangle_{c,v}\right|_{c=1}
  =(\tau\partial_\tau-v\partial_v)\langle Q_\tau^O\rangle_v.
  \label{SMeq: driven_response_operator}
\end{equation}
The driven score is again obtained by differentiating the trajectory likelihood, with the intensity in \cref{SMeq: driven_auxiliary_intensity}.
For the observed record,
\begin{equation}
  \Lambda_{\tau,v}^O
  =\mathbb{E}_v[\Lambda_{\tau,v}^X\,|\,O_\tau],
  \qquad \mathcal{I}_{\tau,v}^O=\langle(\Lambda_{\tau,v}^O)^2\rangle_v.
  \label{SMeq: driven_score_projection}
\end{equation}
Differentiation of the likelihood and the zero mean of the score give
\begin{equation}
  (\tau\partial_\tau-v\partial_v)\langle Q_\tau^O\rangle_v
  =\langle(Q_\tau^O-\langle Q_\tau^O\rangle_v)
           \Lambda_{\tau,v}^O\rangle_v.
  \label{SMeq: driven_score_response}
\end{equation}
The corresponding uncertainty relation is
\begin{equation}
  \frac{[( \tau \partial_\tau - v \partial_v ) \langle Q_\tau^O \rangle_v]^2}{\Var_v[Q_\tau^O]}
  \leq \mathcal{I}_{\tau,v}^O.
  \label{SMeq: driven_uncertainty_relation}
\end{equation}
The full-record score retains the compensated-martingale form in \cref{SMeq: complete_score}, with the history and protocol dependent sensitivity obtained from \cref{SMeq: driven_auxiliary_intensity}.

\section{Time-tagged observables and the waiting-time correction}
\label{SMsec: timestamped_observables}

Consider an autonomous process and an observable that is allowed to depend explicitly on the waiting times,
\begin{equation}
  Q_\tau = Q(\Sigma_\tau, w_0, \ldots, w_{n_\tau}),
  \label{SMeq: timestamped_observable}
\end{equation}
where $w_0 = t_1$, $w_k = t_{k+1} - t_k$ for $1 \leq k < n_\tau$, and $w_{n_\tau} = \tau - t_{n_\tau}$.
Under the change of time variable $s = c t$, a trajectory of the auxiliary process on $[0, \tau]$ corresponds to a trajectory of the original process on $[0, c \tau]$.
Its ordered event sequence is unchanged, but every waiting time in the auxiliary trajectory is $c^{-1}$ times the corresponding waiting time in the original trajectory.
Therefore
\begin{equation}
  \langle Q_\tau \rangle_c =\left\langle Q \! \left( \Sigma_{c\tau}, \left\{ \frac{w_k}{c} \right\}_{k=0}^{n_{c\tau}} \right) \right\rangle,
  \label{SMeq: waiting_time_rescaling_identity}
\end{equation}
where the expectation on the right is taken over the original process up to $c \tau$.
Assuming differentiability in the waiting-time arguments, differentiation at $c = 1$ gives
\begin{equation}
  \left. \partial_{\ln c} \langle Q_\tau \rangle_c\right|_{c=1} = \tau \partial_\tau \langle Q_\tau \rangle -\left\langle \sum_{k=0}^{n_\tau} w_k \partial_{w_k} Q_\tau \right\rangle.
  \label{SMeq: waiting_time_response_correction}
\end{equation}
The second term subtracts the explicit response of the readout to stretching its waiting-time arguments.
It vanishes for observables constructed solely from counts or event order.

The same response can be obtained by differentiation of the full trajectory likelihood,
\begin{equation}
  \left. \partial_{\ln c} \langle Q_\tau \rangle_c \right|_{c=1} = \langle( Q_\tau - \langle Q_\tau \rangle) \Lambda_\tau^X \rangle.
  \label{SMeq: timestamped_score_response}
\end{equation}
Combining \cref{SMeq: waiting_time_response_correction} and \eqref{SMeq: timestamped_score_response}, followed by Cauchy-Schwarz, yields
\begin{equation}
  \frac{ \left[ \tau \partial_\tau \langle Q_\tau \rangle -\left\langle \sum_{k=0}^{n_\tau} w_k \partial_{w_k} Q_\tau \right\rangle \right]^2}{\Var[Q_\tau]} \leq \mathcal{I}_\tau^X.
  \label{SMeq: timestamped_uncertainty_bound}
\end{equation}
This is the full-record inequality quoted in the main text.
Since a time-tagged observable can use information unavailable to $\Sigma_\tau$, its precision need not be bounded by $\mathcal{I}_\tau^\Sigma$; it is bounded by $\mathcal{I}_\tau^X$ with the waiting-time correction shown above.
